\documentclass [prb,preprint,notitlepage,superscriptaddress,floatfix] {revtex4-2}
\usepackage{amsmath}
\usepackage{graphicx}
\usepackage{lmodern}
\usepackage{amsmath}

\usepackage{color}
\usepackage{amssymb}
\usepackage{bm}
\usepackage{braket}
\newcommand{\beq} {\begin{equation}}
\newcommand{\eeq} {\end{equation}}
\newcommand{\bea} {\begin{eqnarray}}
\newcommand{\eea} {\end{eqnarray}}
\newcommand{\be} {\begin{equation}}
\newcommand{\ee} {\end{equation}}

\DeclareMathOperator{\sgn}{sgn}

\usepackage{ulem}
\usepackage[dvipsnames]{xcolor}
\usepackage[colorlinks=true]{hyperref}
\hypersetup{
    colorlinks=true,
    linkcolor=blue,
    citecolor=blue,
    urlcolor=blue
}

\definecolor{darkgreen}{RGB}{0,170,0}

\begin{document}

\title {Odd frequency pairing in a quantum critical metal}
\author{Yi-Ming Wu}
\affiliation{School of Physics and Astronomy and William I. Fine Theoretical Physics Institute,
University of Minnesota, Minneapolis, MN 55455, USA}
\author{Shang-Shun Zhang}
\affiliation{School of Physics and Astronomy and William I. Fine Theoretical Physics Institute,
University of Minnesota, Minneapolis, MN 55455, USA}
\author{Artem Abanov}
\affiliation{Department of Physics, Texas A\&M University, College Station, USA}
\author{Andrey V. Chubukov}
\affiliation{School of Physics and Astronomy and William I. Fine Theoretical Physics Institute,
University of Minnesota, Minneapolis, MN 55455, USA}
\date{\today}

\begin{abstract}
We analyze a possibility for odd-frequency pairing near a quantum critical point(QCP) in a metal.
We consider a model with  dynamical pairing interaction  $V(\Omega_n)\sim 1/|\Omega_n|^\gamma$ (the $\gamma$-model).
 This interaction gives rise to a non-Fermi liquid in the normal state and is attractive for pairing.  The two trends compete with each other.  We search for odd-frequency solutions for the pairing gap $\Delta (\omega_m)= -\Delta(\omega_m)$.  We show that for $\gamma <1$, odd-frequency superconductivity looses the competition with a non-Fermi liquid and does not develop.  We show that the pairing does develop in the extended model in which interaction in the pairing channel is larger than the one in the particle-hole channel.  For $\gamma >1$, we argue that the original model is at the boundary towards odd-frequency pairing and analyze in detail how superconductivity is triggered by a small external perturbation. In addition, we show that for  $\gamma >2$,  the system  gets frozen at the critical point towards pairing in a finite range in the parameter space.  This give rise to highly unconventional phase diagrams with flat regions.
\end{abstract}
\maketitle

\section{Introduction}
\label{sec:introduction}
Superconductivity  in an electronic system appears as a result of a pairing instability, which gives rise
to a formation and subsequent condensation of Cooper pairs. A state with Cooper pairs is described by a complex gap function $\Delta_{\alpha\beta}(\bm{k},\omega)$, where $\alpha, \beta$ are spin indices. The  momentum dependence of $\Delta_{\alpha\beta}(\bm{k},\omega)$ specifies a particular spatial channel $(s-$wave, $p-$wave, $d-$wave, etc) and the frequency dependence determines the behavior of the  spectral function, density of states, and transport properties.

The gap function must satisfy the operational constraint $SP^*T^*=-1$, imposed by fermionic statistics~\cite{Berezinskii,Balatsky1,Balatsky2,Solenov_2009}.
 Here, $S$ is a spin permutation operation, which exchanges spin components $\alpha$ and $\beta$,  $P^*$ is a coordinate permutation operator, which changes $\bm{r}$ into $-\bm{r}$, and $T^*$ is a time permutation operator, which turns $t$ into $-t$.
The application of $SP^*T^*=-1$ sets the condition
\begin{equation}
 	\Delta_{\alpha\beta}(\bm{k},\omega)=-\Delta_{\beta\alpha}(-\bm{k},-\omega)\label{eq:Brule2}
\end{equation}
It allows for two classes of gap functions, as has been
 first noticed by Berezinskii\cite{Berezinskii}: even in frequency gap functions, for which  $\Delta_{\alpha\beta}(\bm{k},\omega)=-\Delta_{\beta\alpha}(-\bm{k},\omega)$ and the ones odd in frequency, for which
$\Delta_{\alpha\beta}(\bm{k},\omega)=\Delta_{\beta\alpha}(-\bm{k},\omega)$.  For even frequency pairing,
 $T^*=1$, hence
 $S P^* =-1$. Then, for spin-singlet pairing ($S=-1$) spatial symmetry must be even
($s$-wave, $d$-wave, etc),  while for spin triplet pairing ($S=1$) it must be odd ($p$-wave, $f$-wave, etc.)

For odd-frequency pairing, the gap function is odd under time permutation, $T^*=-1$,  and the identity $SP^* T^* =-1$ requires that  for spin singlet pairing spatial symmetry must be $p$-wave, $f$-wave, etc., while  for spin triplet pairing it must be $s$-wave, $d$-wave, etc.

We emphasize that neither even-frequency nor odd-frequency pairing breaks time-reversal symmetry as under the action of anti-unitary time reversal operator ${\hat T} = e^{-i\pi {\hat S}_y} {\hat K}$, where ${\hat K}$ imposes complex conjugation, we have
${\hat T} \Delta_{\alpha\beta}(\bm{k},\omega){\hat T}^{-1} = \Delta^*_{\alpha\beta}(\bm{k},-\omega)$. For even frequency pairing ${\hat T} \Delta_{\alpha\beta}(\bm{k},\omega){\hat T}^{-1}$ can be restored back to $\Delta_{\alpha\beta}(\bm{k},\omega)$ by changing the phase of $\Delta_{\alpha\beta}(\bm{k},\omega) = |\Delta_{\alpha\beta}(\bm{k}, \omega)| e^{i\phi}$ from $\phi$ to $-\phi$, while for odd-frequency pairing the corresponding change is $\phi \to \pi-\phi$.

Odd-frequency superconductivity is a rare phenomenon, yet there have been multiple efforts  to detect it in nature, particularly in disordered superconductors~\cite{Kirkpatrick_1991, Zuzin_2019},  heterostructures, and external driven fields\cite{Linder1,Houzet,Linder2015,Shigeta_2012}.
 One idea here is to take an even-frequency superconductor
  and put it in contact with an external source, which breaks time-reversal symmetry and  creates
    a ``field'' for an odd-frequency gap component.  This was  proposed to develop
      near the interface between a conventional $s-$wave superconductor and a ferromagnet~\cite{Bergeret1,Volkov1} and for an s-wave superconductor in a magnetic field, before FFLO state sets in~\cite{matsumoto2012coexistence,matsumoto2012coexistence_1,tsvelik2019superconductor}.
 Another idea is to induce an odd-frequency gap component near the interface between a
triplet superconductor and a normal metal due to the breakdown of even/odd rule under
coordinate permutation near the interface~\cite{Tanaka1,Tanaka2,Tanaka3,Tanaka4,Eschrig1,Eschrig2,Buzdin}.  Yet another idea is to
 combine disorder and a closeness to an ordinary $s-$wave superconductivity and explore fluctuation-induced pre-emptive instability towards an odd-frequency pairing~\cite{Zuzin_2019}.
 Besides,  odd-frequency pairing was argued to develop in the vicinity of ordered states with broken time-reversal or translation symmetry~\cite{Hakansson2015}

 A spontaneous development of an odd-frequency superconductivity without an external ``field'' requires an attraction in the odd-frequency channel, but even if this is the case, one has to  overcome three obstacles.
 First, because an odd-frequency gap vanishes at zero frequency, there is no Cooper logarithm, and hence a non-zero solution of the gap equation can emerge only if the coupling is strong enough.  Second, if the coupling is strong, fermionic self-energy is also strong, and it acts against pairing by reducing the magnitude of the pairing kernel.  Third, in many cases, an attraction in the odd-frequency channel is accompanied by a similar-in-strength attraction in the even-frequency channel. An even-frequency pairing then develops at a higher $T$
   because of Cooper logarithm and, once developed,  acts against odd-frequency pairing by again reducing the
     magnitude of the kernel for off-frequency pairing.

 The first two obstacles are quite generic and one needs to analyze specific models to see whether they can be overcomed.
 In particular, it was  argued that the destructive effect from the self-energy can be reduced if
  the irreducible interaction in the pairing channel is stronger than the one in the particle-hole channel
 ~\cite{Abrahams_1993,Abrahams_1993_E,fuseya2003realization,pimenov_4}.
  The third obstacle can potentially be eliminated by
  bringing the system to the end point of an even-frequency pairing by varying the strength of
 an instant Hubbard repulsion, which  negatively affects even-frequency pairing but does not influence odd-frequency paring channel. Along these lines, it was conjectured that  an odd-frequency pairing may develop near the end point of even-frequency $p-$wave superconductivity~\cite{matsumoto2013emergent} and near the end point of phonon-mediated s-wave superconductivity~\cite{pimenov_3}.

In multiband/muliorbital systems, there is an additional band/orbital index,  which has to be treated on equal footings with the spin index. it was argued~\cite{Black-Schaffer,Aperis,Asano} that a hybridization between different bands/orbitals can lead to a realization of an odd-frequency pairing. Finally,
a somewhat different pairing state, also termed as
  odd-frequency
  superconductivity, has been proposed to develop in the Kondo-lattice model, due to the process involving three-body scattering~\cite{Tsvelik_1,Tsvelik_2,Tsvelik_3}, and in the Kondo-Heisenberg model~\cite{Tsvelik_4,Tsvelik_5}.

   In this paper, we analyze odd-frequency pairing for a set of quantum-critical  systems, in which the pairing is mediated by a critical gapless  boson
  ~\cite{acf,acs,acs2,sslee,sslee2,Subir,Subir2,moon_2,max,max2,raghu,raghu2,raghu3,raghu4,raghu5,review,review2,review3,
  review4,max_last,wang,CW,mack,steve_sam,berg,berg_2,berg_3,kotliar,kotliar2}.  Such an interaction is strongly frequency dependent, and the gap equation allows, at least in principle,  both even-frequency and odd-frequency solutions.
    We specifically consider a set of critical systems, in which an effective dynamical 4-fermion interaction $V(\Omega_m)$,  channeled into a proper spatial channel, scales as
      $V(\Omega_m) \propto 1/|\Omega_m|^\gamma$  (the $\gamma$-model).
  For the discussion of the application of this model to various fermionic systems see, e.g., Ref. \cite{paper1}. Even-frequency, spin-singlet pairing in the $\gamma-$model has been analyzed in several recent publications~\cite{Wang1,Wu2,Chubukov1,paper_1,paper_2,paper_3,paper_4,paper_5,paper_6}. Here, we analyze odd-frequency pairing in the spin-triplet channel for the same set of models and, in each case, for the same spatial symmetry.
  For definiteness, we assume that even-frequency pairing is eliminated by, e.g., strong frequency-independent repulsive component of the interaction, and consider how an odd-frequency pairing potentially emerges due to the exchange of a gapless dynamical boson, thus circumventing the 3rd problem mention above.

  We argue that in the canonical $\gamma$-model with the same interaction  $V(\Omega_m)$ in the particle-hole and  particle-particle channels, odd-frequency pairing does not develop because  fermionic self-energy keeps the attractive pairing interaction below the threshold. For electron-phonon interaction (the case $\gamma =2$) this has been obtained previously  for a finite Debye frequency (see Ref.~\cite{Balatsky2} and references therein).   Our results show that this holds even when a pairing boson becomes massless.
   However, odd-frequency pairing does develop in the model with different interactions in the two channels,
   if
  the one in the pairing channel is larger. For $\gamma <1$, this happens when ratio of the two interactions  exceeds a certain threshold. For $\gamma >1$, the pairing develops immediately once the pairing interaction exceeds the one in the particle-hole channel.  A recent study of vertex corrections to Elishberg theory for the $\gamma =2$ model with electron-phonon attraction and Hubbard repulsion did find~\cite{pimenov_4} that the dressed  interaction in the particle-particle  is larger than the dressed interaction in the particle-hole channel.

Below we express our results in terms of  $D (\omega) = \Delta (\omega)/\omega$, which is an even function of frequency, and in many respects is the analog of $\Delta  (\omega)$ for even-frequency pairing.

 We study odd-frequency pairing in the $\gamma$-model separately for  $\gamma<1$ and $\gamma >1$.
   For $\gamma <1$, we model non-equivalence of the interactions in  particle-particle and  particle-hole channels
    by multiplying the pairing interaction by the factor $1/N$ and treating $N$  as a parameter, smaller than one.
  We find the critical $N_{cr} (\gamma) <1$ that separates a non-Fermi liquid ground state at $N > N_{cr}$ and a superconducting ground state at $N < N_{cr}$ (more accurately, a state with a non-zero pairing gap).

   We argue that $N_{cr}$ is a multi-critical point , below which there emerges an infinite discrete set of topologically different odd-frequency functions $D_n (\omega)$ with $n=0,1,2,...$.  The magnitude of $D_n (\omega \to 0)$ is the largest for $n=0$ and at large enough $n$ decreases as $e^{-A n}$, $A = O(1)$.
    A topological distinction between $D_n (\omega)$ with different $n$ shows most clearly on the Matsubara axis, where  $D_n (\omega_m)$ can be made real by a proper  choice of an overall phase.   The function $D_n (\omega_m)$   has $n$  nodes at finite positive $\omega_m$, and the equal  number of nodes at negative $\omega_m$.  Each node of $D_n(\omega_m)$ is a center of a $2\pi$ vortex
    in the complex frequency plane,
     hence the $D_n (\omega_m)$ has $n$ vortices on the positive part of the Matsubara axis.

  As a proof that the infinite set of $D_n (\omega_m)$ does exist, we  obtain the exact analytical solution
     at $T=0$ for the end-point of the set, $D_{n=\infty}$.  We also obtain numerically the onset temperatures $T_{p,n}$  for the gap functions with $0 \leq n \leq 8$.  We show  that these $T_{p,n}$  are non-zero and decay exponentially with $n$, like $D_n (0)$ at $T=0$.
 We show that all $T_{p,n}$ vanish at $N =N_{cr}$
  \footnote{This is similar, but not identical to the even-frequency pairing, where $T_{p,n}$ with $n \geq 1$ vanish at $N = N_{cr}$, but $T_{p,0}$ only vanishes at $N \to \infty$. The reason for this behavior  is a special role of fermions with the first
  Matsubara frequencies $\pm \pi T$ for even-frequency pairing. We show that for odd-frequency pairing fermions with $\omega_m = \pm \pi T$ are not special. As a consequence, $T_{p,0}$ vanishes at the same $N= N_{cr}$ as other $T_{p,n}$.}.
 We emphasize that an infinite set of $D_n (\omega_m)$ exists only for pairing at a QCP, when a pairing boson is massless.
  For odd-frequency pairing out of a Fermi liquid away from a QCP, the number of solutions becomes finite. The number of solutions depends on the distance to a QCP, and above a certain distance only $D_0 (\omega_m)$ survives, i.e., the gap equation has a single solution.

   We next analyze the form of $D_0 (\omega_m)$
   at $T=0$.
   We solve the non-linear gap equation and analyze how  $D_0 (\omega_m)$
   depends on $\omega_m$. For odd-frequency pairing out of a
   Fermi liquid, $\Delta (\omega_m) \propto \omega_m$ at small $\omega_m$, hence
   $D_0 (\omega_m)$ tends to a constant.  We argue that at a QCP the low-frequency behavior is different: $D_0 (\omega_m)$ diverges as $1/|\omega_m|^{d}$, where $d$ depends on $\gamma<1$ and satisfies $0<d<1$. On the real frequency axis, this leads to a non-analytic density of states (DoS)  at small frequencies: $N(\omega) \propto \omega^{d}$. To obtain $N(\omega)$ at arbitrary $\omega$, we analytically continue $D_0 (\omega_m)$  onto the real axis using Pade approximants. We obtain a complex $D (\omega) = D' (\omega) + i D^{''} (\omega)$ and show that  $D^{'} (\omega)$ passes through zero at a frequency where $D'' (\omega) \approx 1$.
This gives rise to a sharp peak in $N(\omega)$, reminiscent of edge singularity  for even-frequency pairing.
We note in passing that  there is no zero-bias peak in our case, in distinction to some models of odd-frequency pairing out of a Fermi liquid in S/N and S/F heterostructures~\cite{Tanaka4,DiBernardo2015}.

We then extend the analysis to  $\gamma >1$. We argue that  for these $\gamma$, $N_{cr} =1$, i.e., the canonical model with equal interaction in particle-particle and particle-hole channels is critical for odd-frequency pairing. We show that the pairing emerges for arbitrary $N<1$, and the onset temperature for the pairing scales as $T_p \sim (1/N-1)^{1/(\gamma-1)}$. We find that, like the case of $\gamma<1$, there exists an infinite number of solutions, of which the onset temperatures $T_{p,n}$ are not identical, although scale in the same way with $1/N-1$. Then, for each $\gamma >1$, there is a universal tower of onset temperatures for pairing in different topological sectors, specified by $n$.
 We show that a similar behavior holds at $T=0$, if we keep bosonic mass $\omega_D$ finite: there is
   a set of crirical $\omega_D$, all of order $ (1-N)^{1/(\gamma-1)}$, but with different $n-$dependent prefactors.

We show that the scaling relations $T_{p,n} \propto (1/N-1)^{1/(\gamma-1)}$ and  $\omega_{D,n} \propto (1/N-1)^{1/(\gamma-1)}$ emerge because for $\gamma >1$ and $N <1$, the gap equation contains infrared singularities,  and a finite $T$ or a finite $\omega_D$ act to regularize these singularities.   We also discuss another extension of the model with $\gamma >1$ to non-equal interactions in the pairing and the particle-hole channel,  which does not induce infrared singularities.  Using this extension, characterized by the parameter $M$, we find $M_{cr} (\gamma)$, at which odd-frequency superconductivity emerges.  We show that it emerges simultaneously for all $n \geq 0$ for $1< \gamma <2$, but a new physics emerges for $\gamma >2$, and, as a result,  the order with $n=0$ emerges prior to ordering in other topological sectors.

 The paper is organized as follows. In Sec.~\ref{sub:Justification} we consider two most frequently cited examples
of pairing at a QCP - pairing by nematic and antiferromagnetic fluctuations in 2D.
We show that in both cases there is an attraction in the odd-frequency channel, and the  gap equation is
formally
identical to
 the one for even-frequency pairing. In Sec.~\ref{sec:model} we introduce a generic $\gamma-$model for odd-frequency pairing at a QCP and  extend it to $N \neq 1$. In Sec.~\ref{sec:solutions_on_matsubara_axis}, we consider the range $0 <\gamma<1$. In Sec.~\ref{sub:linearized_equation_critical_}-\ref{sub:nonlinear_gap_equation}, we analyze the linearized and the nonlinear gap equation on the Matsubara axis and establish the condition for odd-frequency pairing. We show that these exists a set of topologically distinct solutions $\Delta_n (\omega)$, each with its own onset temperature $T_{p,n}$. We obtain the exact solution at $T=0$ for $n =\infty$ and discuss in some length the solution at $n=0$, for which the condensation energy is the largest. In Sec.~\ref{sec:solutions_on_real_axis} we analytically continue the gap function with $n=0$  to the real frequency axis. We obtain a complex $D_0 (\omega)$ and use it to obtain the DoS  $N(\omega)$. In Sec.\ref{sec:the_case_} we extend the analysis to $\gamma>1$. In Sec.~\ref{subsubsec:M_b0_finiteT}, we discuss how the solutions with different $n$ appear one-by-one once we consider the limit when $1-N$ and $T$  (or bosonic mass $\omega_D$) are both vanishingly small, but the ratio $(1/N-1)/T^{\gamma-1}$ stays finite. In Sec.~\ref{sec:sol2} we discuss hidden physics at $\gamma >2$. To unravel it we extend the model with $\gamma >1$  to non-equal interactions in the particle-particle and particle-hole channels without introducing infra-red
 singularities.
 We present our conclusions in Sec.~\ref{sec:conclusion_and_discussion}.

\section{Examples of odd-frequency pairing at a QCP}
\label{sub:Justification}
In this section we analyze the two most known examples of pairing near a QCP in a 2D metal -- pairing by Ising-nematic charge fluctuations and by antiferromagnetic spin fluctuations. We adopt the same strategy as for  even-frequency pairing:
  introduce the pairing vertex $\Phi(\omega_m,\bm{k})$ and the fermion self energy $\Sigma(\omega_m,\bm{k})$, and assume that a critical boson is overdamped and is  slow  compared to a fermion near a Fermi surface. This approximation allows one to understand the pairing and its competition with non-Fermi liquid by analyzing the set of two coupled Eliashberg equations for $\Phi(\omega_m,\bm{k})$ and  $\Sigma(\omega_m,\bm{k})$.  After momentum integration, which can be done explicitly, the set reduces to two coupled equations for $\Phi (\omega_m)$ and $\Sigma (\omega_m)$.  We show that the equations have the same form for even-frequency and odd-frequency pairing, but for odd-frequency pairing $\Phi (\omega_m)$ obeys $\Phi (-\omega_m) =- \Phi (\omega_m)$.

\subsection{Pairing at a 2D Ising-nematic QCP}
\label{subsub:pairing_at_a_nematic_qcp}
The susceptibility of an Ising-nematic order parameter is peaked at momentum $q=0$, and its low-energy dynamics is determined by Landau damping into particle-hole pairs. The
 effective 4-fermion interaction, mediated by a massless boson at a charge-nematic QCP, is
\begin{equation}
	V_{\text{eff}} (\bm{q},\Omega_m)=\frac{g_{\text{eff}}}{\bm{q}^2+\Gamma\frac{|\Omega_m|}{|\bm{q}|}}\label{eq:chi}
\end{equation}
where $g_{\text{eff}}$ is fermion-boson coupling and  $\Gamma=g_{\text{eff}} m/(\pi v_F)$  (for a parabolic dispersion of fermions, $\xi_k = (k^2-k^2_F)/(2m)$).
As we said, we assume that a Landau overdamped  boson is a slow mode compared to a low-energy fermion.  One can verify that this holds when $g_{\text{eff}} \ll E_F$.
 In this situation, one can
 approximate $V_{\text{eff}}$ by its value for $q =2k_F \sin {\theta/2}$, connecting points ${\bf k}$ and ${\bf p}$ on the Fermi surface (${\bf p} = {\bf k} + {\bf q}$, ${\bf p} {\bf k} = k^2_F \cos{\theta}$).
 In a lattice system, $\Gamma$ depends on the angle between ${\bf q}$ and a particular direction in the Brillouin zone, but this dependence does not change the results qualitatively and we proceed assuming a rotational invariance.

The Eliashberg equation for spin-singlet pairing vertex is
 \begin{equation}
  \Phi(\omega_n,\theta_k)= \pi T\sum_{\omega_m}\int\frac{pdpd\theta_p}{(2\pi)^2}\frac{\Phi(\omega_m,\theta_p)
  }{\xi_p^2+\tilde{\Sigma}^2(\omega_m)+\Phi^2(\omega_m,\theta_p)} V_{\text{eff}} (\theta_p-\theta_k,\omega_m-\omega_n)\label{eq:Phi_angle}
\end{equation}
Here $\tilde{\Sigma} (\omega_m) = \omega_m + \Sigma (\omega_m)$,  $\theta_k$ and $\theta_p$ are the angles with respect to some arbitrary chosen direction in the Brillouin zone, and $V_{\text{eff}} (\theta_p-\theta_k,\omega_m-\omega_n)$ is obtained by substituting $\bm{q}=2k_F\sin(\frac{\theta_p-\theta_k}{2})$ into Eq.\eqref{eq:chi}.  We will assume and then verify that relevant values of $\omega_n$ and $\omega_m$ are of order $g_{\text{eff}} \ll E_F$. Typical $|\theta_k - \theta_p|$ are then parametrically small in $g_{\text{eff}}/E_F$.  To leading order in $g_{\text{eff}}/E_F$, one can then approximate $\Phi^2(\omega_m,\theta_p)$ in the r.h.s. of (\ref{eq:Phi_angle}) by $\Phi^2(\omega_m,\theta_k)$ and explicitly integrate over $\theta_k-\theta_p$ and by $\xi_p$. In doing this, we do not distinguish between even- and odd-frequency pairing.
Performing the integration, we obtain $0+1$ dimensional equation for $\Phi (\omega_n, \theta_k)$:
\begin{equation}
		\Phi(\omega_n, \theta_k)=\pi T  \sum_{\omega_m}\frac{\Phi(\omega_m, \theta_k)}{\sqrt{\tilde{\Sigma}^2(\omega_m)+\Phi^2(\omega_m,\theta_k)}}\left(\frac{\bar{g}}{|\omega_n-\omega_m|}\right)^{1/3}\label{eq:nematic}
\end{equation}
 where $\bar{g}  = \frac{1}{162\sqrt{3}\pi^2}
  g_{\text{eff}}^2/E_F$.
   We see that $\Phi (\omega_n, \theta_k)$ doesn't actually depend on $\theta_k$, hence the gap equation does not distinguish between even-frequency spin-singlet pairing, for which $\Phi$ is  even under $\theta_k \to \pi + \theta_k$,
and odd-frequency spin-singlet pairing,
  for which
  $\Phi$ is odd under $\theta_k \to \pi + \theta_k$.   An alternative way to see this is to divide the Fermi surface into patches and verify that
 fermions from different patches don't talk to each other~\cite{Lee_review}.

\begin{figure}
  \includegraphics[width=8.5cm]{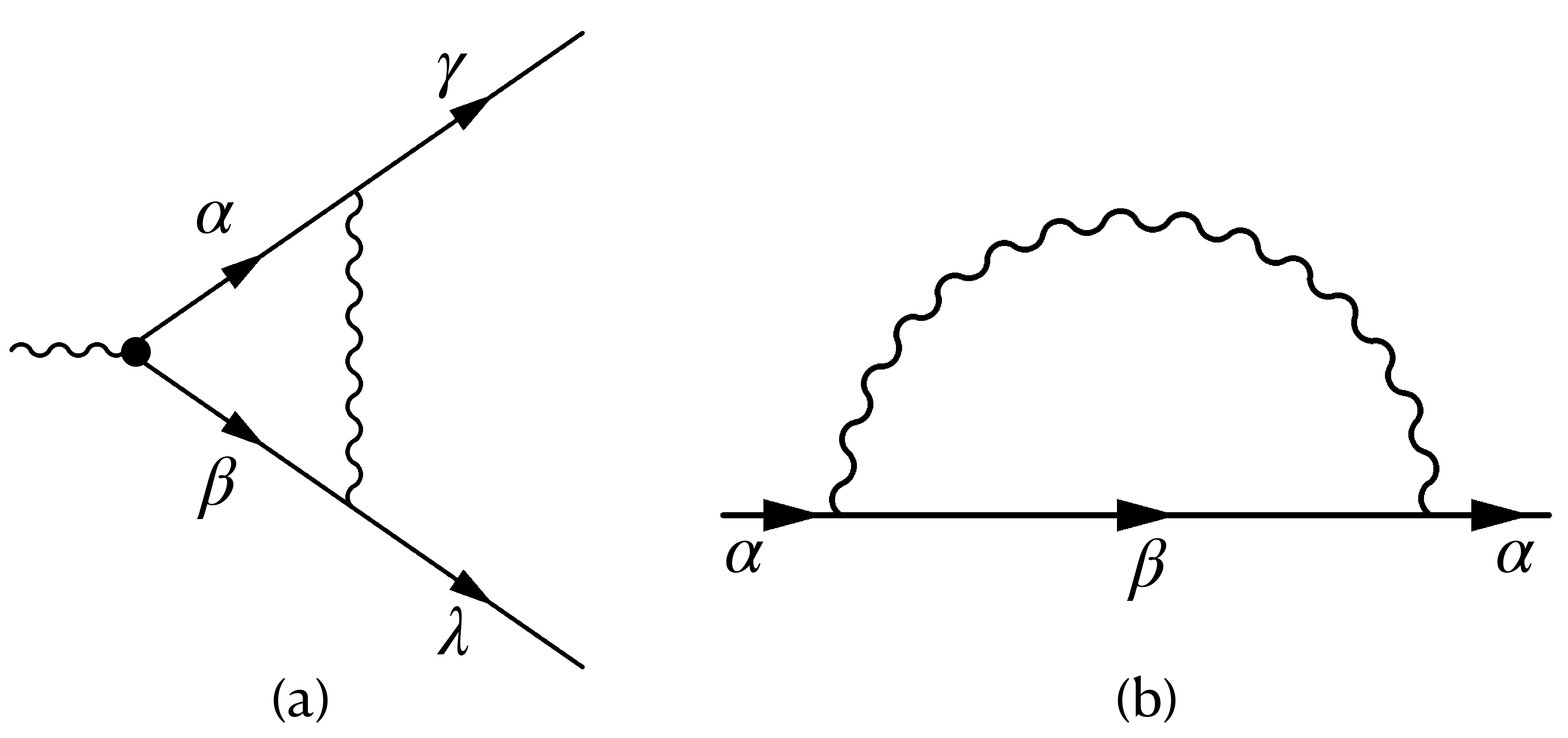}
  \caption{(a) Pairing vertex with spin-spin interaction. $\alpha,\beta,\gamma,\lambda$ are spin indices. For spin-singlet pairing we have $i\sigma^{y}_{\alpha\beta}\bm{\sigma}_{\alpha\gamma}\cdot\bm{\sigma}_{\beta\lambda}=-3i\sigma^y_{\gamma\lambda}$ while for spin-triplet pairing we have $(\sigma^i i\sigma^{y})_{\alpha\beta}\bm{\sigma}_{\alpha\gamma}\cdot\bm{\sigma}_{\beta\lambda}=(\sigma^i i\sigma^{y})_{\gamma\lambda}$. (b) Self-energy with spin-spin interaction. The extra factor of $3$ can be seen from contracting the internal spin index $\sum_\beta\bm{\sigma}_{\alpha\beta}\cdot\bm{\sigma}_{\beta\alpha}=(\bm{\sigma}^2)_{\alpha\alpha}=3$.}\label{fig:spin}
\end{figure}

 For the interaction mediated by Ising spin fluctuations at a spin-nematic QCP,  the gap equation is again the same for even- and odd-frequency
 pairing vertices. The only difference with the charge case is that now there is an  extra factor of $3$ for the self-energy and the overall factor of either $-3$ or $1$ for spin-singlet and spin-triplet pairing (see Fig. \ref{fig:spin}).   A non-zero $\Phi (\omega_n, \theta_k)$ is then only possible for the spin-triplet pairing.

 \subsection{Pairing at a 2D anti-ferromagnetic QCP}
\label{subsub:pairing_at_a_2d_anti_ferromagnetic_qcp}
The pairing interaction, mediated by soft antiferromagnetic  fluctuations, is peaked at momentum  $\bm{Q}=(\pi,\pi)$, and its dynamics again comes from Landau damping:
\begin{equation}
	V^{sf}_{\alpha \beta;\gamma \delta} (\bm{q},\Omega_m) = V^{sf} (\bm{q},\Omega_m)  {\bf \sigma}_{\alpha \gamma}{\bf \sigma}_{\beta,\delta},~~V^{sf}  (\bm{q},\Omega_m) = \frac{g_{\text{eff}}}{(\bm{q}-\bm{Q})^2+\Gamma_{sf}|\Omega_m|}
\end{equation}
\begin{figure*}
	\includegraphics[width=15cm]{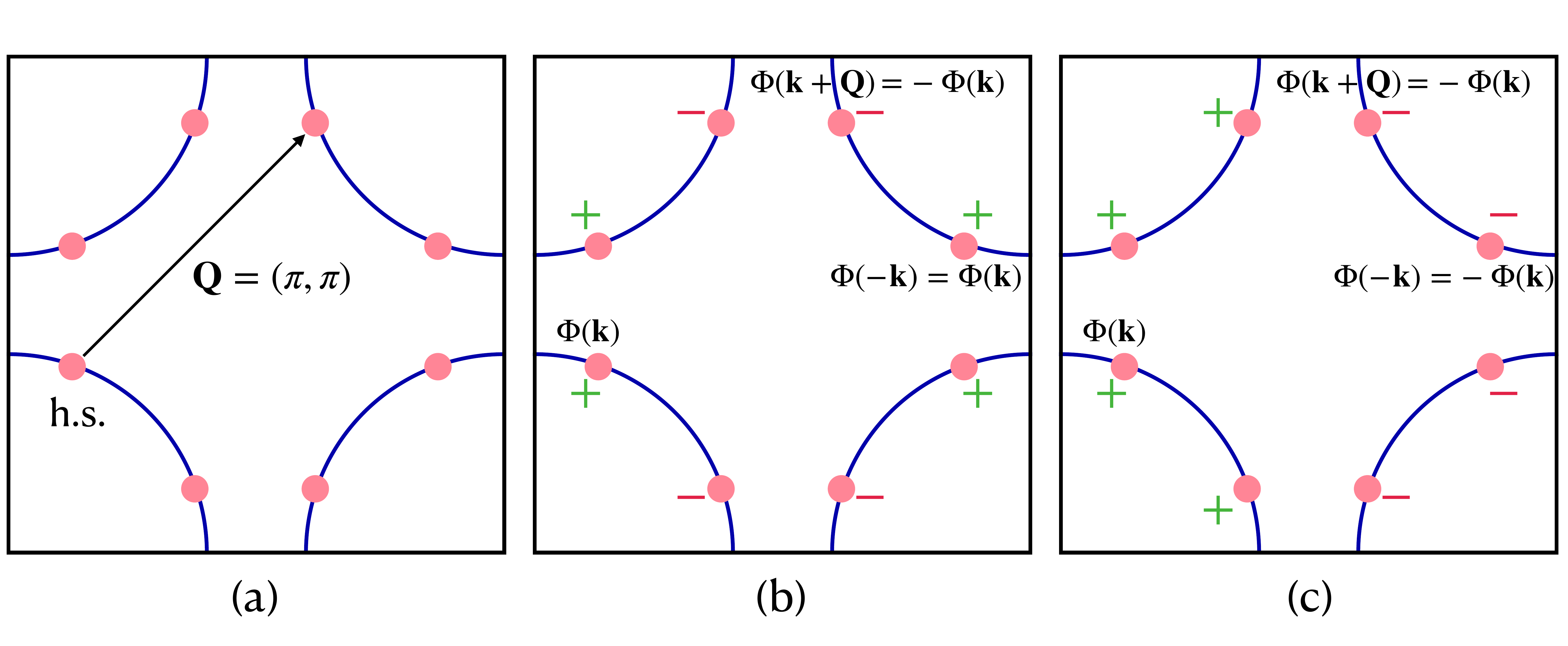}
	\caption{(a) Fermi Surface of hot spot model. When the interaction is peaked at $\bm{Q}=(\pi,\pi)$, there are 8 hot spots on the Fermi surface which can be pair-connected by momentum $\bm{Q}$ either directly or via Umklapp. (b) For spin-singlet even-frequency paring, we must have $\Phi(\omega_m,-\bm{k})=\Phi(\omega_m,\bm{k})$, thus the pairing symmetry is most likely d-wave. (c) For spin-singlet odd-frequency pairing, we must have $\Phi(\omega_m,-\bm{k})=-\Phi(\omega_m,\bm{k})$. In this situation, the pairing symmetry is obviously p-wave.}\label{fig:fs}
\end{figure*}
The gap equation for spin-singlet pairing is
\begin{equation}
\Phi(\omega_n,\bm{k})=-3T\sum_{\omega_m}
\int\frac{d^2\bm{p}}{(2\pi)^2}\frac{\Phi(\omega_m,\bm{p})}{\xi_{p}^2+\tilde{\Sigma}^2(\omega_m)+\Phi^2(\omega_m,\bm{p})}
V_{sf}(\bm{k}-\bm{p},\omega_n-\omega_m)\label{eq:sf}
\end{equation}
The factor `$-3$' originates from spin summation (Fig.\ref{fig:spin}).
 Its presence implies that an $s-$wave solution is impossible.

Because $(\pi,\pi)$ is a lattice wave vector, one has to consider lattice dispersion and a non-circular Fermi surface.  Motivated by the cuprates, we consider the Fermi surface, shown in Fig.\ref{fig:fs}(a). The momentum $\bm{Q}$
 connects 8 points on this Fermi surface (hot spots) $\bm{k}_{h.s.}$ either directly or via Umklapp.  For $g_{eff} \ll E_F$,
  one can safely neglect fermions located away from hot regions and focus on fermions in the patches around hot spots.

 To overcome the overall minus sign in the r.h.s. of (\ref{eq:sf}), we search for $\Phi(\omega_m,\bm{k})$ which satisfy the condition $\Phi(\omega_m,\bm{k}_{h.s.}+\bm{Q})=-\Phi(\omega_m,\bm{k}_{h.s.})$.
 The difference between even- and odd-frequency pairing in this situation is in the parity of $\Phi(\omega_m,\bm{k})$. For even-frequency pairing, we must have $\Phi(\omega_m,-\bm{k})=\Phi(\omega_m,\bm{k})$, while for odd-frequency pairing, the Berezinskii rule imposes the condition $\Phi(\omega_m,-\bm{k})=-\Phi(\omega_m,\bm{k})$.
  A simple experimentation shows that in the even-frequency case, $\Phi(\omega_m,\bm{k})$  has $d_{x^2-y^2}$ symmetry, with nodes along Brillouin zone diagonals (Fig.\ref{fig:fs}(b)), while in the odd-frequency case, $\Phi(\omega_m,\bm{k})$  has p-wave symmetry (Fig.\ref{fig:fs}(c)).  For both cases,  one can set $k = k_{h.s.}$ in Eq. (\ref{eq:sf}), approximate $\Phi(\omega_m,\bm{p})$ by $\Phi(\omega_m,\bm{k}_{h.s.}+\bm{Q})=-\Phi(\omega_m,\bm{k}_{h.s.})$, and explicitly integrate
  over the two components of momenta:
    over the one transverse to the Fermi surface in the kernel and over the one along the Fermi surface in the interaction.
      This yields
\begin{equation}
		\Phi(\omega_n)=\pi T\sum_{\omega_m}\frac{\Phi(\omega_m)}{\sqrt{\tilde{\Sigma}^2(\omega_m)+\Phi^2(\omega_m)}}
\left(\frac{\bar{g}}{|\omega_n-\omega_m|}\right)^{1/2}\label{eq:AFM}
\end{equation}
where $\bar{g}=\frac{9}{16\pi} g^2_{\text{eff}}/(\Gamma_{sf}v_F^2)$
 and $\Phi(\omega_n) = \Phi(\omega_n, {\bf k}_{h.s.})$.
We emphasize that this equation holds  for both  even- and odd-frequency pairing.  To differentiate  between the two, one
 has to move away from hot regions and solve for the gap on the full Fermi surface.  For $g_{\text{eff}} \ll E_F$,
   the pairing vertex  along the full Fermi surface is induced by $\Phi(\omega_n, {\bf k}_{h.s.})$, and for both even-frequency and odd-frequency pairing the onset pairing temperature, obtained from the full Fermi surface
     analysis, differs little from the one obtained by solving Eq. (\ref{eq:AFM}).

The same analysis can be performed for other cases of pairing at a QCP with the same result -- after the momentum integration the equation for the dynamical pairing vertex $\Phi (\omega_m)$  has the same form for even-frequency and odd-frequency pairing.

\section{Model}
\label{sec:model}

We analyze odd-frequency pairing using the same strategy as for even-frequency one. Namely, we consider  an itinerant fermion system close to a QCP towards
 charge or spin ordering and assume that the dominant interaction between fermions is mediated by soft fluctuations of
   an order parameter that condenses at the transition.  The resulting effective 4-fermion interaction is attractive for odd-frequency pairing.   The same interaction, however,
    gives rise to a singular self-energy, which leads to  incoherent  non-Fermi liquid (NFL) behavior in the normal state.
 These two tendencies compete in the sense that a NFL behavior in the normal state reduces the pairing kernel, while once fermion pair,  low-energy excitations become gapped, and the  self-energy recovers a Fermi liquid form.

 We assume, like in earlier studies (see~\cite{paper_1} and references therein),
 that
 order parameter fluctuations are slow modes compared to fermions, and that there exists
  a small parameter, which allows one to neglect vertex  corrections.
      In this situation one can select the most attractive spatial pairing channel and explicitly integrate over momentum along and transverse to the Fermi surface. After this,
      the problem  reduces to the analysis of $0+1$ dimensional coupled dynamical integral equations for the pairing vertex $\Phi (\omega)$ and
    fermionic self-energy  $\Sigma (\omega)$.
   On the Matsubara axis, these equations are (we use $\tilde{\Sigma} (\omega_{n} )\equiv \omega_{n}+\Sigma (\omega_{n})$) \bea
	&&\Phi(\omega_m)=\pi T\sum_{n}\frac{\Phi(\omega_n)}{\sqrt{\tilde{\Sigma}(\omega_n)^2+\Phi(\omega_n)^2}}
 V\left(|\omega_n-\omega_m|\right) \label{eq:Eliahsberg1_1}\\
&&		\tilde{\Sigma}(\omega_m)=\omega_m+\pi T\sum_{n }\frac{\tilde{\Sigma}(\omega_n)}{\sqrt{\tilde{\Sigma}(\omega_n)^2+\Phi(\omega_n)^2}}
V\left(|\omega_n-\omega_m|\right)
	\label{eq:Eliahsberg1}
\eea
 where $V\left(|\omega_n-\omega_m|\right) = V(\Omega_m)$ is the effective local dynamical interaction, taken for fermions on the Fermi surface and integrated over momentum transfer along the Fermi surface. Eqs.~(\ref{eq:Eliahsberg1_1}-\ref{eq:Eliahsberg1}) are similar to the Eliashberg equations for electron-phonon interaction, and we will be calling them Eliashberg equations. We refer to Ref.~\cite{paper1} for the discussion of the model and the justification of the Eliashberg-type theory.

The pairing gap $\Delta (\omega_m)$ is related to the pairing vertex as
\beq
\Delta (\omega_m) = \Phi (\omega_m) \frac{\omega_m}{\omega_m + \Sigma (\omega_m)}
\eeq
We emphasize that the two equations (\ref{eq:Eliahsberg1_1}-\ref{eq:Eliahsberg1}) have the same form for even-frequency and odd-frequency pairing. Below we focus on odd-frequency solution for $\Phi (\omega_m)$.

In a Fermi liquid, $V(\Omega_m)$ is a constant at small $\Omega_m$.  There is no solution for odd-frequency pairing for a near-constant $V(\Omega_m)$, hence the system remains in the normal state. As the system approaches a QCP,
$V(\Omega_m)$ acquires progressively stronger dependence on $\Omega_m$.  This allows one to search for odd-frequency  solutions $\Phi (-\omega_m) = - \Phi (\omega_m)$.

At a QCP,  $V(\Omega_m)$  can be generally written  as $V(\Omega_m)=(\bar{g}/|\Omega_m|)^\gamma$, where $\bar{g}$ is an effective coupling constant with dimension of energy, and $\gamma>0$ is an exponent.
  The two examples, considered in the previous section, correspond to  $\gamma=1/3$ and $\gamma=1/2$, respectively.

 The two  Eliashberg equations can be  re-arranged into the equation for the pairing gap $\Delta (\omega_m)$ and the quasiparticle residue $Z(\omega_m) = {\tilde \Sigma} (\omega_m)/\omega_m = 1 + \Sigma (\omega_m)/\omega_m$ :
\bea
	&&  \Delta(\omega_m)=\pi T\sum_{\omega_n}\frac{\Delta(\omega_n)-\Delta(\omega_m)\frac{\omega_n}{\omega_m}}{\sqrt{\omega_n^2+\Delta^2(\omega_n)}}
 \left(\frac{\bar{g}}{|\omega_m-\omega_n|}\right)^\gamma\label{eq:Delta} \\
 && Z (\omega_m) = 1 +
 \frac{\pi T}{\omega_m} \sum_{\omega_m} \frac{\omega_n}{\sqrt{\omega^2_n +\Delta(\omega_n)^2}}
 \left(\frac{\bar{g}}{|\omega_m-\omega_n|}\right)^\gamma
 \label{eq:Z}
\eea

Observe that  the numerator in (\ref{eq:Delta}) vanishes at $n=m$.  This holds even if the bosonic propagator has a finite mass.
Treating  a QCP as the limit when a bosonic mass vanishes, we  can then safely eliminate the term with $m=n$ from the r.h.s. of (\ref{eq:Delta}). Physically this implies that thermal fluctuations do not affect the gap equation, similar to non-magnetic impurities
in the even frequency $s$-wave case.
This doesn't hold for $Z(\omega_m)$, which at a finite $T$ does contain a singular thermal contribution from  $\omega_m = \omega_n$.  This term has to be properly regularized.  The same is true for  $\Phi$ and $\Sigma$, which do require regularization. Below we focus on the solution of the equation for the gap function $\Delta (\omega_m)$.  This equation does not require a regularization  for $\gamma <3$, which we  only consider.

\subsection*{Extension to \texorpdfstring{$N\neq1$}{}}
\label{sub:extension_to_}

The set of Eliashberg Eqs.~(\ref{eq:Eliahsberg1_1}-\ref{eq:Eliahsberg1}) assumes that exactly the same effective dynamical interaction $V(\Omega_m)$ appears in the particle-particle and the particle-hole channel.  Previous works on even-frequency pairing have shown that one can get better understanding of  the interplay between pairing and non Fermi liquid by analyzing  a generalized $\gamma-$model with non-equal interactions in the two channels.  This can be rigorously done by extending the original $U(1)$ model to matrix $SU(N)$ (Ref. \cite{raghu_15}). For such a model, the Eliashberg equation for the self-energy remains intact, but in the one for the pairing vertex there appears a factor $1/N$  in the r.h.s..  We follow earlier works on even-frequency pairing~(see e,g., Ref.\cite{paper_1} and references therein) treat $N$  as a continuous parameter, which measures relative strength of the interactions in the particle-particle and the particle-hole channel.  If $N >1$, the interaction in the  particle-hole channel is stronger, while  if $N <1$, the one in the particle-particle channel is stronger.

  As we said, our primary goal is the analysis of the gap equation. The extension to $N \neq 1$ changes this equation to
\begin{equation}
	\Delta(\omega_m)=\pi T\sum_{\omega_n}\frac{\frac{1}{N}\Delta(\omega_n)-\Delta(\omega_m)\frac{\omega_n}{\omega_m}}
{\sqrt{\omega_n^2+\Delta^2(\omega_n)}}\left(\frac{\bar{g}}{|\omega_m-\omega_n|}\right)^\gamma\label{eq:Delta_N}
\end{equation}
The extension has to be done carefully to avoid generating a singular thermal contribution.   The way to do this is to first eliminate the $n=m$ term in the gap equation (\ref{eq:Delta}) and only then extend the model to  $N \neq 1$.

For odd-frequency pairing it is convenient to re-express Eq. (\ref{eq:Delta_N}) in terms of $D(\omega_m) = \Delta (\omega_m)/\omega_m$, which in this case is an even function of frequency and in this respect is similar to the gap function for even-frequency pairing.
 The equation for $D(\omega_m)$ is
 \begin{equation}
	D(\omega_m) \omega_m=\pi T\sum_{n \neq m }\frac{\left(\frac{1}{N} D(\omega_n)-D(\omega_m)\right) {\text{sign}}  (\omega_n)}
{\sqrt{1 +D^2(\omega_n)}}\left(\frac{\bar{g}}{|\omega_m-\omega_n|}\right)^\gamma\label{eq:Delta_N1}
\end{equation}
At $T=0$, the frequency summation is replaced by the integration:
 \begin{equation}
	D(\omega_m) \omega_m=\frac{1}{2} \int d \omega_n \frac{\left(\frac{1}{N} D(\omega_n)-D(\omega_m)\right) {\text{sign}}  (\omega_n)}
{\sqrt{1 +D^2(\omega_n)}}\left(\frac{\bar{g}}{|\omega_m-\omega_n|}\right)^\gamma
\label{eq:Delta_N1T0}
\end{equation}

 To understand whether an odd-frequency pairing emerges  above a QCP,
 one needs to analyze the linearized  gap equation
 \begin{equation}
  D(\omega_m) \omega_m=\pi T\sum_{n \neq m} \left(\frac{1}{N}D(\omega_n)-D(\omega_m)\right) {\text{sign}} (\omega_n) \left(\frac{\bar{g}}{|\omega_m-\omega_n|}\right)^\gamma\label{eq:Delta_N2}
\end{equation}
and check whether it acquires a non-zero solution at some $T =T_p$.

To simplify the notations, we will be calling the equation for $D(\omega_m)$ the gap equation.
  we will also measure, $T$, $\omega_m$, and $\Delta$ in units of ${\bar g}$.

\section{Solutions on Matsubara axis, \texorpdfstring{$\gamma <1$}{}.}
\label{sec:solutions_on_matsubara_axis}

\subsection{linearized equation at \texorpdfstring{$T=0$: $N_{cr}$}{}}
\label{sub:linearized_equation_critical_}

We first analyze the linearized gap equation at zero temperature.
 The goal here is to find whether there exists a critical $N_{cr}$ separating a non-Fermi liquid ground state and a superconducting ground state.
At $T=0$, the gap equation  in \eqref{eq:Delta_N2} becomes
\begin{equation}
	D(\omega_m)\omega_m=\frac{1}{2}\int_{-\infty}^\infty d\omega_n\left(\frac{1}{N}D(\omega_n)-D(\omega_m)\right)\frac{\text{sign}(\omega_n)}{|\omega_m-\omega_n|^\gamma}
\label{eq:linearT0}
\end{equation}

  We use the same strategy as for even-frequency pairing: focus on low frequencies and search for power-law solution for $D(\omega_m)$ in the form $D(\omega_m) \propto |\omega_m|^{\alpha}$, where $\alpha >-1$.
If $\alpha$ is real, $D(\omega_m)$ is sign-preserving and  gradually evolves from the bare infinitesimally small $D_{bare} (\omega_m)$, as one can easily verify. The pairing susceptibility $\chi_{pp} = D(\omega_m)/D_{bare} (\omega_m)$ then remains finite, hence there is no pairing instability. If, however, the exponent $\alpha$ is complex, there must be two
 complex conjugated exponents
 $\alpha = \alpha' \pm i\alpha^{''}$.
 The real function $D(\omega_m)$ then oscillates at
     small frequencies as
     $D(\omega_m) \sim  |\omega_m|^{\alpha'} \cos\left(\alpha^{''} \log{\omega_m} + \phi \right)$, where $\phi$ is some number.
     An oscillating behavior is incompatible with an iterative expansion staring from $D_{bare} (\omega_m)$, as iterations retain  $D(\omega_m)$ sign-preserving, and implies that the system is unstable towards pairing.

\begin{figure}
	\includegraphics[width=8.5cm]{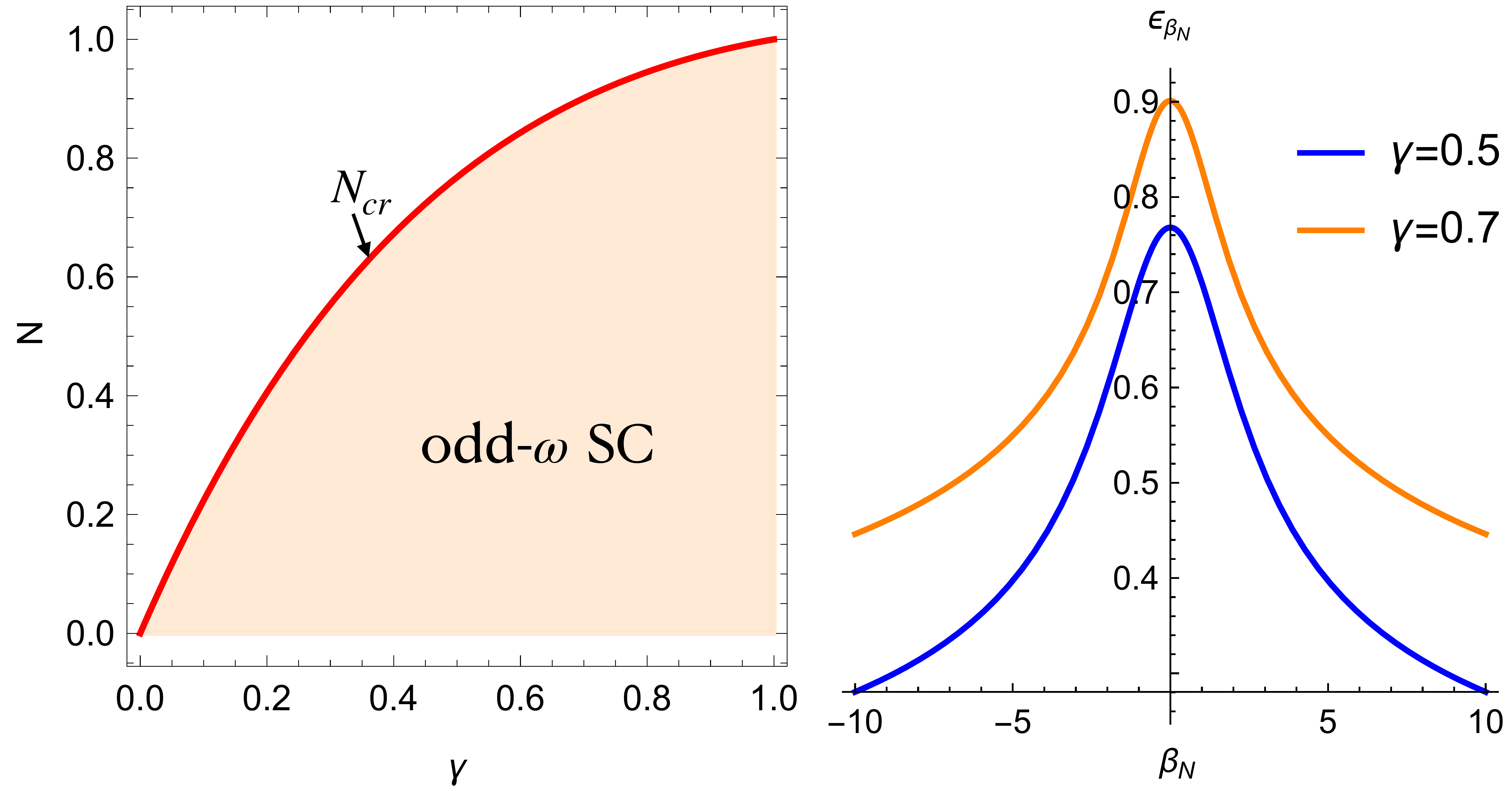}
	\caption{(a)$N_{cr}$ as a function of $\gamma$ for odd-frequency pairing. The pairing develops for $N < N_{cr}$. (b) The function $\epsilon_{\beta_N}$ in \eqref{eq:beta} for two representative values of $\gamma$. The equation $N=\epsilon_{\beta_N}$ has solutions when $N<N_{cr}$, and $N_{cr}$ is the maximum value of $\epsilon_{\beta_N}$ taken at $\beta_N=0$ ($N_{cr}=0.77$ for $\gamma=0.5$ and $N_{cr}=0.90$ for $\gamma=0.7$.).}\label{fig:Ncr}
\end{figure}

 Substituting $D(\omega_m) \propto |\omega_m|^{\alpha}$ into (\ref{eq:linearT0}) we find that at small frequencies the
   $D(\omega_m) \omega_m$ term in the l.h.s. of (\ref{eq:linearT0}) can be neglected, and the gap equation reduces to
  \begin{equation}
	\frac{1}{2N}\int_{-\infty}^\infty d\omega_n |\omega_n|^\alpha
\frac{\text {sign}(\omega_n)}{|\omega_m-\omega_n|^\gamma}= |\omega_m|^\alpha
\int_{-\infty}^\infty d\omega_n \frac{\text {sign} (\omega_n)}{|\omega_m-\omega_n|^\gamma}
\label{eq:linearT0_1}
\end{equation}
 Solving for $\alpha$ we find that it is real when $N > N_{cr}$ and complex when $N < N_{cr}$, where
\begin{equation}
   N_{cr} = \frac{1-\gamma}{2\Gamma(\gamma)}|\Gamma(\gamma/2)|^2\left(\frac{1}{\cos(\pi \gamma/2)}-1\right)	\label{eq:beta_0}
\end{equation}
    The complex exponent for $N < N_{cr}$ is
 $\alpha =  \gamma/2-1  \pm i \gamma \beta_N$,
   where $\beta_N$ is the solution of
  \begin{eqnarray}
 N &=& \epsilon_{\beta_N}, \nonumber \\
 \epsilon_{\beta_N}&=&
\frac{1-\gamma}{2\Gamma(\gamma)}|\Gamma(\gamma/2+i\gamma\beta_N)|^2\left(\frac{\cosh(\pi\gamma\beta_N)}{\cos(\pi \gamma/2)}-1\right)	
\label{eq:beta}
\end{eqnarray}
 We plot $N_{cr} (\gamma)$ in Fig.\ref{fig:Ncr}(a). We see that $N_{cr}$ exists, but is smaller than one for all $\gamma$ from the range $0 <\gamma <1$.
 This implies that the canonical $\gamma$-model with equal interactions in the particle-hole and particle-particle channels remains stable against the odd frequency pairing down to $T=0$.
 However, when the pairing interaction gets larger, the system eventually becomes unstable towards pairing.
 In Fig.\ref{fig:Ncr}(b) we plot $\epsilon_{\beta_N}$ as a function of $\beta_N$ for two representative values of $\gamma$. We see that $N=\epsilon_{\beta_N}$ has two solutions $\pm\beta_N$ when $N<N_{cr}$, and $\beta_N$ increases as $N$ decreases below $N_{cr}$. For $N \leq N_{cr}$, $\beta_N \propto (N_{cr} -N)^{1/2}$.

 On a more careful look we note that for $N > N_{cr}$, there are actually two power-law solutions of (\ref{eq:linearT0_1}) with real exponents $\alpha_1$ and $\alpha_2$.  At large $N$ when the interaction in the pairing channel is weak, $\alpha_1  \approx \gamma$ comes from integration over internal $\omega_n \gg \omega_m$ (an UV contribution), while the other $\alpha_2 \approx2$ comes from integration over $\omega_n \ll \omega_m$ (an IR contribution). The latter is not connected to iterations starting from $D_{bare} (\omega_m)$ and from this perspective is  an unphysical solution. As $N$ decreases, $\alpha_1$ and $\alpha_2$ move towards each other: $\alpha_{1,2} = \gamma/2-1 \pm b_N$, where $b_N$ decreases from its value $1-\gamma/2$ at large $N$.
  At $N = N_{cr}$, $b_N$ vanishes, and the two solutions merge.  At smaller $N$, $b_N$ becomes imaginary ($i\gamma\beta_N$), and $\alpha_{1,2}$ become complex. Right at $N=N_{cr}$, a more accurate analysis of (\ref{eq:linearT0_1}) shows that
  there are again two solutions: $D_1(\omega_m) = |\omega_m|^{\gamma/2-1}$ and $D_1(\omega_m) = |\omega_m|^{\gamma/2-1} \log{|\omega_m|}$. Just like for even-frequency pairing, combining the two solutions  at small frequencies into $D(\omega_m) = D_1 (\omega_m) + c D_2 (\omega_m)$ and using $c$ as a parameter, one can match the low-frequency form of $D(\omega_m)$ with the high-frequency form and obtain the solution of the linearized gap equation  for all $\omega_m$,
   as it should be the case at the onset of superconductivity.   The large frequency form of $D(\omega_m)$ is
   $D(\omega_m) \propto 1/|\omega_m|^{\gamma +2}$, as one straightforwardly extract from (\ref{eq:linearT0}).

    The linearized gap equation  at $N = N_{cr}$  can actually be solved exactly using the same approach as Ref. \cite{paper_1} (see more on this below).
      The function $D(\omega_m)$ is sign-preserving and interpolates between two different power-law forms at small and large  frequencies.

 \subsection{linearized equation at a finite \texorpdfstring{$T$}{} }

\begin{figure}
    \includegraphics[width=7.5cm]{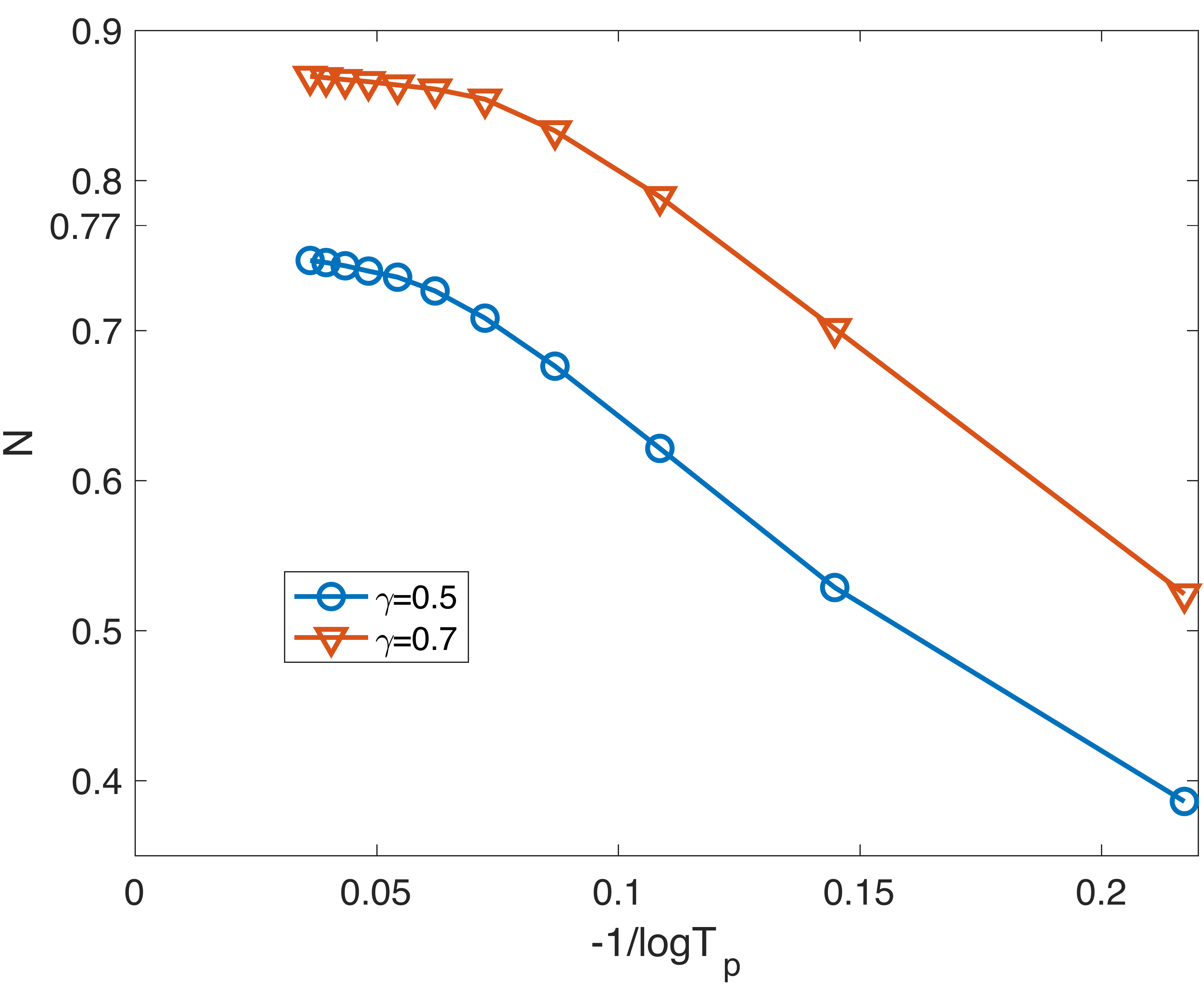}
    \caption{$T_p(N)$ from solving the linearized equation \eqref{eq:Delta_N2} for $\gamma=0.5$ and $0.7$. For better illustration, we plot $-1/\log T_p$ instead of $T_p$ on the horizontal axis. Both curves clearly extrapolate to their corresponding $N_{cr}$ in zero temperature limit.}\label{fig:Tp0}
\end{figure}

 We solved the linearized gap equation at a finite $T$ numerically, focusing on the sign-preserving solution. We  find the critical temperature $T_p (N)$, at which the solution of the linearized equation exists. We show $T_p (N)$  in Fig.\ref{fig:Tp0}
obtained by numerically solving the linearzied equation \eqref{eq:Delta_N2}. We see that $T_p (N)$ is non-zero for $N < N_{cr}$ and terminates at $N = N_{cr}$. This behavior is intuitively expected, yet we emphasize that it differs from that for even-frequency pairing. There, $T_p (N)$ by-passes $N_{cr}$ and terminates at $N =\infty$.  The difference is in the gap equation for the lowest Matsubara frequencies $\omega_m = \pm \pi T$. This contribution is generally a special one because the contribution to the gap equation from the fermionic self-energy (the term with external $D(\omega_m)$ in the r.h.s. of (\ref{eq:linearT0})) vanishes for $\omega_m =\pm \pi T$ as  $ \sum_{n \neq 0} {\text{sign}} (2n+1) / |2 \pi T n|^\gamma =0$. Hence, for $\omega_m = \pm \pi T$, there is no competitor to pairing.  For even-frequency pairing, the interaction between fermions with $\omega_m = \pi T$ and $\omega_m=-\pi T$ is attractive, and solving for $T_p$ one obtains $T_p = (1/2\pi) (1/N)^{1/\gamma}$, i.e., $T_p$ vanishes at $N = \infty$ rather than at $N = N_{cr}$. This creates a special re-entrance behavior of the gap function, which emerges at $T_p$ but
   then vanishes at $T=0$.  For odd-frequency pairing, the interaction between fermions with $\omega_m = \pi T$ and $\omega_m=-\pi T$ is repulsive as $\Delta (-\pi T) = - \Delta (\pi T)$. Because of repulsion,  fermions with
$\omega_m = \pm \pi T$
 are not  relevant for the pairing.  The latter comes from fermions with other $\omega_m$, for which self-energy is finite and at a small but  finite $T$ is of the same order as at $T=0$.
     In this situation, the line  $T_c (N)$ terminates at $N = N_{cr}$

\begin{figure}
	\includegraphics[width=8cm]{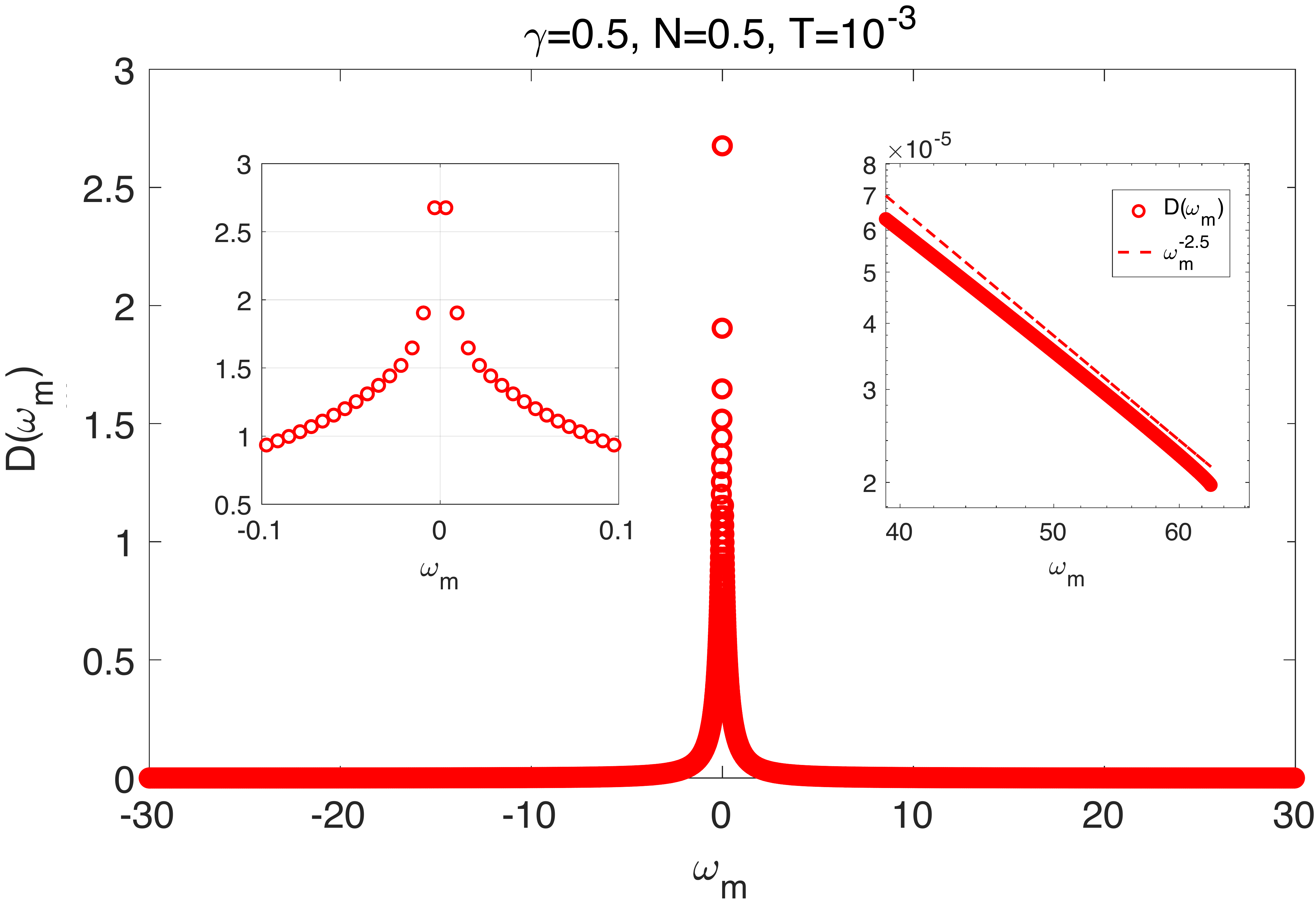}
	\caption{$D(\omega_m)$ obtained  from the numerical solution of the nonlinear gap equation for $\gamma=0.5$, $N=0.5$ and $T=10^{-3}$.  The insets show small and large $\omega_m$ behaviors of $D(\omega_m)$.
 }\label{fig:oddDelta}
\end{figure}

\subsection{Nonlinear gap equation}
\label{sub:nonlinear_gap_equation}

The full non-linear gap equation (\ref{eq:Delta_N1}) can be solved by numerical iterations.
  The  gap function  emerges at $T_p$ and reaches maximum value at $T=0$.  We show $D(\omega_m)$ at small
  $T \sim 10^{-3} {\bar g}$
  for  representative $\gamma =1/2$ and $N = 0.5 < N_{cr}$ in Fig.~\ref{fig:oddDelta}. This gap function is sign-preserving, which implies that the oscillations in the pairing susceptibility are all eliminated by a finite $D(\omega_m)$
   Below we analyze this gap function at $T=0$ both analytically and numerically, at the lowest possible temperature.
   We verified that for such $D(\omega_m)$ the iteration procedure is fully convergent.

\begin{figure}
	\includegraphics[width=8cm]{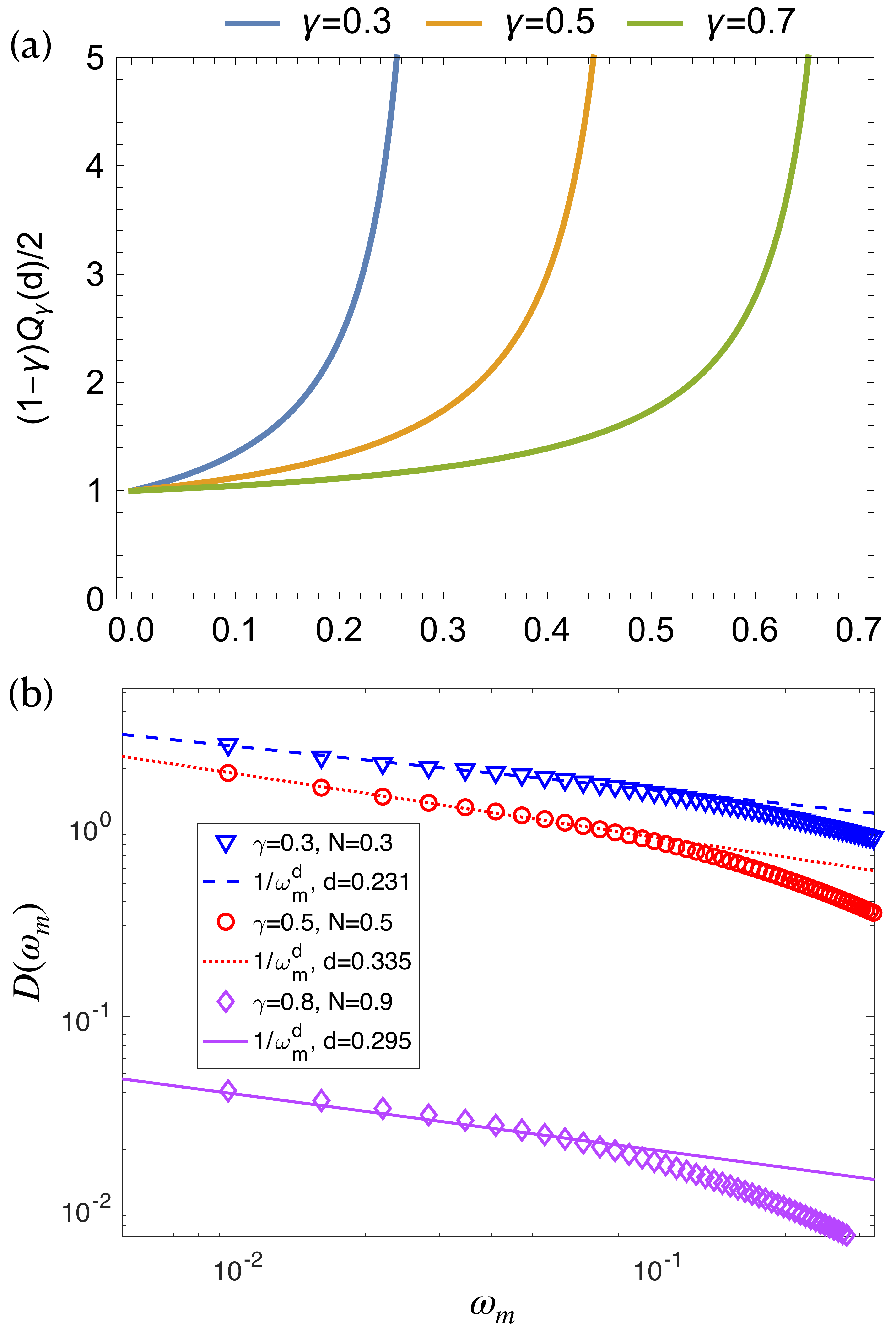}
	\caption{(a) The function $F_\gamma(d)=(1-\gamma)Q_\gamma(d)/2$ for different $\gamma$ and $0<d <\gamma$.
 In the two limits, $F_\gamma(0) =1$ and $F_\gamma (\gamma)$ diverges. In between, $F_\gamma (d)$  monotonically increases with $d$.  This behavior guarantees that the equation $F_\gamma (d) = 1/N$ has a solution for any $N < N_{cr} <1$.
    (b) Comparison between $D(\omega_m) \sim 1/|\omega_m|^d$ and the numerical solution of the full nonlinear gap equation at small frequencies.  The agreement is perfect, but holds only for low enough frequencies.
    }\label{fig:Fa}
\end{figure}

  At large frequencies $D(\omega)$  scales as $1/|\omega_m|^{5/2}$, as is expected ($5/2= 2 +\gamma$ for $\gamma =0.5$).  At small frequencies,
 one would naively expect that odd-frequency $\Delta (\omega_m)$ should be linear in $\omega_m$ and hence  $D(\omega_m)$ should approach a finite value at $\omega_m \to 0$.
 Fig.~\ref{fig:oddDelta} however shows that $D(\omega_m)$ is actually non-analytic and diverges at $\omega_m \to 0$.
 This singular behavior  can be understood analytically. Indeed, $D (\omega_m) \approx {\text {const}}$ at $\omega_m \to 0$  does not satisfy the gap equation (\ref{eq:Delta_N1}) at $T=0$, as  both sides of this equation contain $\omega_m$ as the overall factor, and the prefactor in the l.h.s. if finite, but the one   in the r.h.s.
diverges as $\int d \omega_n/|\omega_n|^{\gamma +1}$.  Searching for a singular $D(\omega_m) \propto 1/|\omega_m|^d$ with $0<d <1$, we find that the gap equation is satisfied if
\begin{equation}
\frac{(1-\gamma)}{2}Q_\gamma(d)=\frac{1}{N}\label{eq:fa}
\end{equation}
 where
\begin{equation}
	\begin{aligned}
		Q_\gamma(d)=&\int_0^\infty dx x^{d}\left(\frac{1}{|x-1|^\gamma}-\frac{1}{(x+1)^\gamma}\right)\\
		=&B(d+1,1-\gamma)+B(1-\gamma,\gamma-1-d)\\
    &-B(d+1,\gamma-1-d)\\
	\end{aligned}
\end{equation}
where $B(p,q)$ is the Beta function. We plot $(1-\gamma) Q_\gamma (d)$ for different $\gamma$ in Fig.\ref{fig:Fa}(a). This function is equal to one for $d=0$, increases with $d$ and diverges at $d =\gamma$. This guarantees that  Eq. (\ref{eq:fa}) has a solution
  at  some $0<d<\gamma$  for any $N < N_{cr} <1$. We also numerically confirmed the small-$\omega_m$ scaling of $D(\omega_m)$ in Fig.\ref{fig:Fa}(b), via a direct comparison between the numerical solution of the nonlinar equation and $1/\omega_m^d$ with $d$ determined from \eqref{eq:fa}.

\subsection{Gap function  on the real axis}
\label{sec:solutions_on_real_axis}

To obtain  spectral observables, like the fermionic density of states (DOS) or the spectral function, we need to know the
 function $D(\omega)$ on the real axis.  We obtained it using two procedures: (i) Pade approximants method and  (ii) by transforming the gap equation to the real axis and solving it there.  In the second procedure we followed the same steps as for even frequency pairing. We obtain the same results for $D(\omega)$ in both ways.  Below we show the results obtained by Pade approximants.  The quality of this method is tested by
  first obtaining $D(\omega)$ on the real axis from $D(\omega_m)$ and then re-evaluating $D(\omega_m)$ by Cauchy  formula and comparing with the original  $D(\omega_m)$.  The relative difference between the two is
  at most $10^{-7}$.
\begin{figure}
	\includegraphics[width=8.5cm]{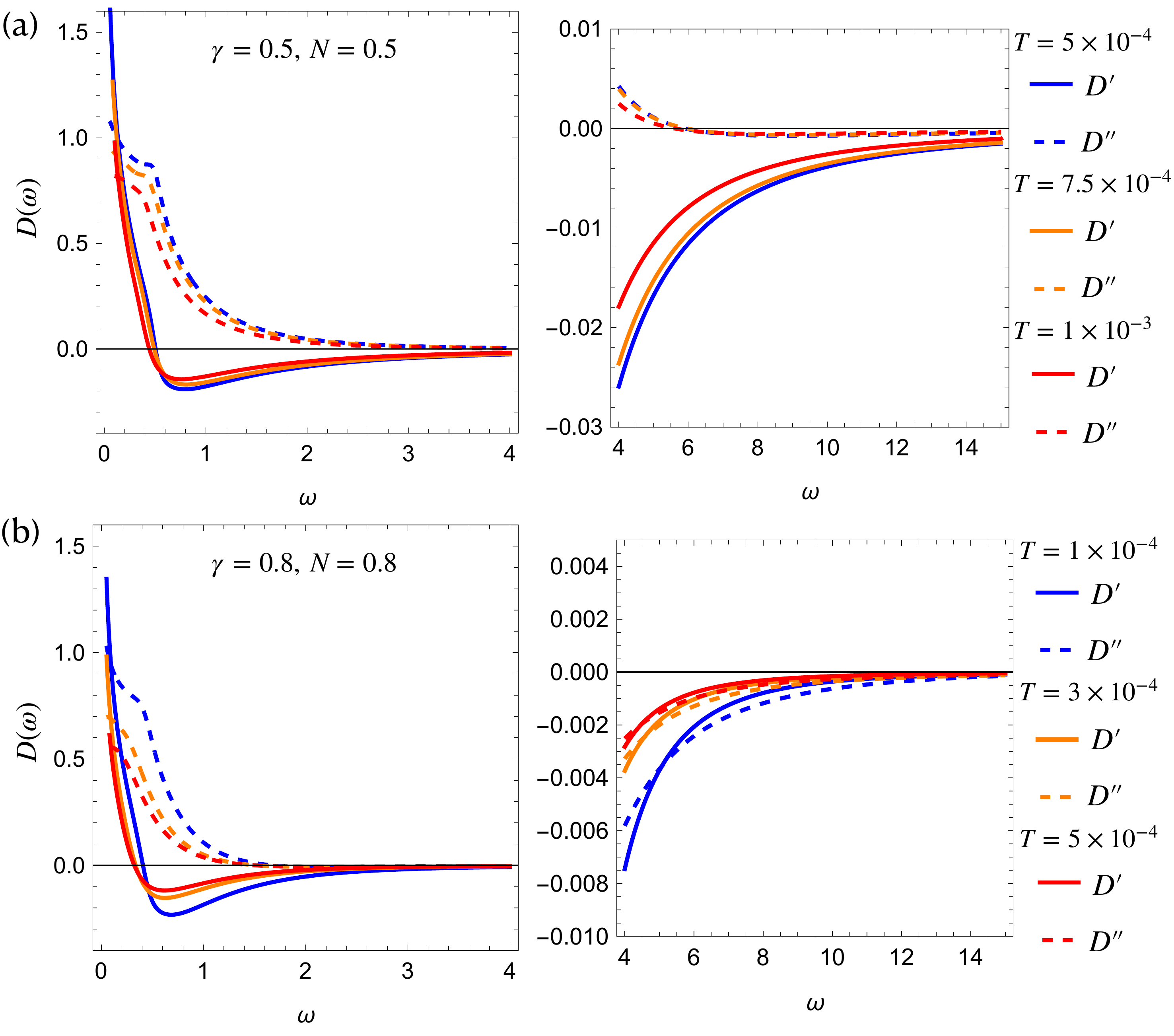}
	\caption{The gap function $D(\omega)=D'(\omega)+iD''(\omega)$ along the real axis  for $\gamma=0.5$, $N=0.5$ and $\gamma=0.8$, $N=0.8$, at different temperatures. Left panels and right panels show the behavior at small and large $\omega$, respectively. }\label{fig:realDelta}
\end{figure}

In Fig.~\ref{fig:realDelta} we show the numerical results for $D (\omega) = D^{\prime} (\omega) + i D^{\prime\prime} (\omega)$ for
two values of $(\gamma,N)$. We recall that we set $D(\omega_m)=\Delta(\omega_m)/\omega_m$ to be real.
  By Cauchy relation we then have on the real axis $D^{\prime}(-\omega) =  D^{\prime}(\omega)$ and $D^{\prime\prime} (-\omega) = -D^{\prime\prime}(\omega)$, i.e., $D^{\prime}(\omega)$ is even and $D^{\prime\prime} (\omega)$ is odd. This is similar to the case of even-frequency pairing, where for real $\Delta (\omega_m)$, $\Delta^{\prime} (\omega)$ is even and $\Delta^{\prime\prime} (\omega)$ is odd. We also note in passing that there a qualitative difference between odd-frequency superconductivity and
even-frequency gapless superconductivity. For the latter,  the pairing gap vanishes at $\omega=0$, but is even in frequency.
Then $D(\omega_m)$ is odd,  and by  Cauchy relation on the real axis  $D^{\prime} (\omega)$ is odd and $D^{\prime\prime} (\omega)$ is even.

  Back to our case. At small and large frequencies, a simple rotation of the Matsubara axis expression yields $D^{\prime} (\omega) = \cos(\pi d/2)/|\omega|^d$ and $D^{\prime\prime} (\omega) = \sgn(\omega)\sin(\pi d/2)/|\omega|^d$.  At large frequencies, $D^{\prime} (\omega) = -\cos(\pi \gamma/2)/|\omega|^{2+\gamma}$ and $D^{\prime\prime} (\omega) = -\sgn(\omega)\sin(\pi \gamma/2)/|\omega|^{2+\gamma}$  ($d = 0.335$ for $(\gamma,N)=(0.5,0.5)$ and $d=0.45$
   for $(\gamma,N)=(0.8,0.8)$.)  Comparing these two forms, we immediately find that both $D^{\prime}$ and $D^{\prime\prime}$ have to change sign between small and large $\omega$. These small and large frequency behaviors can be clearly seen in Fig.~\ref{fig:realDelta}.

 We use $D(\omega)$ on the real axis to obtain the DOS
\begin{equation}
  N(\omega)=\text{Re}\left(\frac{1}{\sqrt{1+D(\omega) D^* (-\omega)}}\right)
\label{eq:N}
\end{equation}
 To the best of our knowledge, this expression was first obtained in Ref.\cite{Solenov_2009}.

\begin{figure}
	\includegraphics[width=8.5cm]{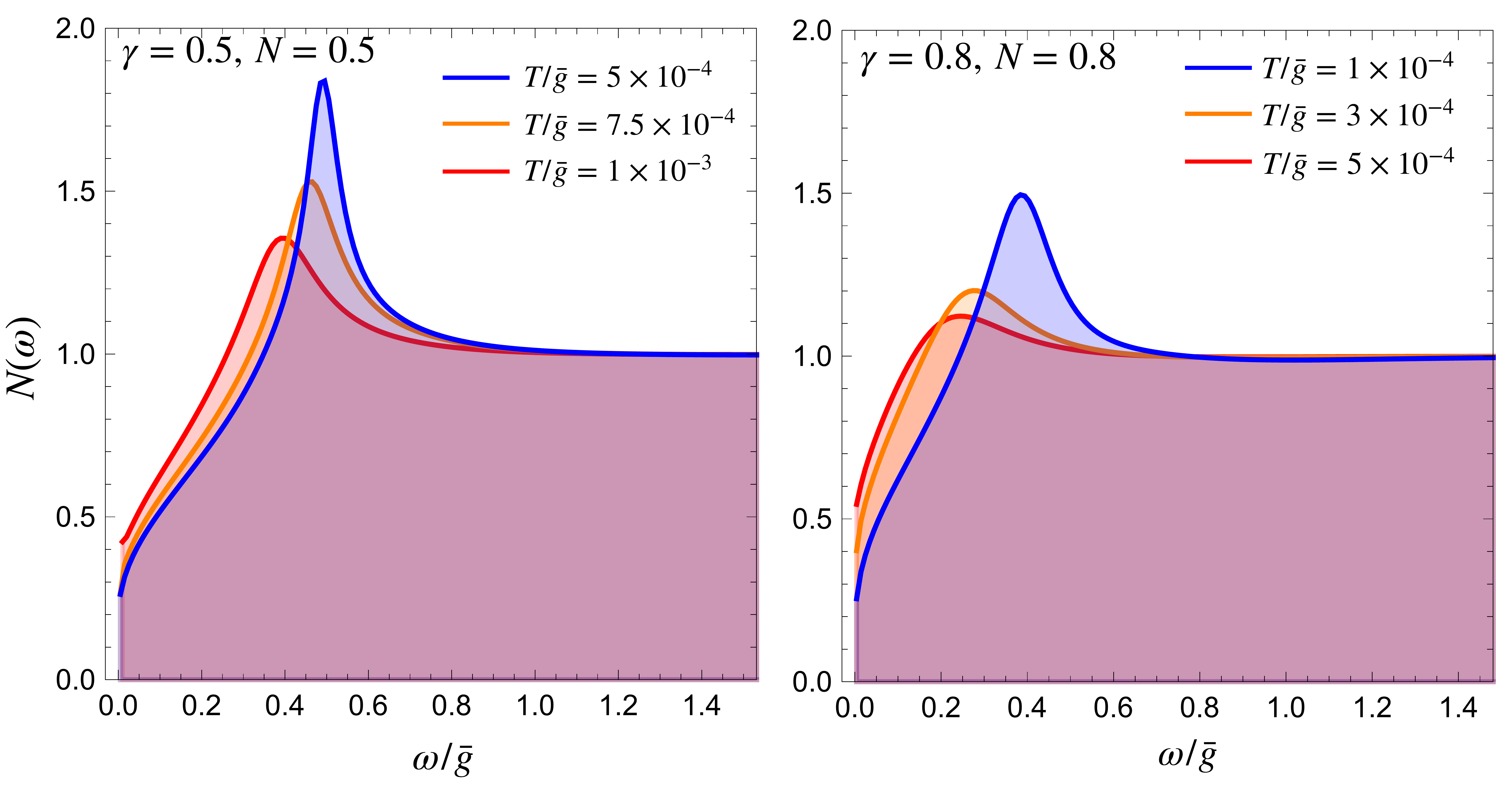}
	\caption{The DOS $N(\omega)$, normalized to its normal state value $N_0$, at various temperatures for $(\gamma, N)=(0.5,0.5)$ and $(0.8,0.8)$.  }\label{fig:DOS_1}
\end{figure}

  We show $N(\omega)$ in Fig.\ref{fig:DOS_1} for the two different $(\gamma, N)$ from Fig.\ref{fig:realDelta}.
   In both cases, we see that the DOS is strongly reduced at small $\omega$ and has a has a sharp peak at $\omega \leq 1$.
    Both features can be understood analytically.  The reduction of $N(\omega)$ at small $\omega$ is the  consequence of the fact that at $T=0$,  both $D'$ and $D''$ diverge at $\omega \to 0$ as $1/|\omega|^d$, hence $N(\omega) \propto |\omega|^d$. The peak at $\omega \leq 1$ comes about because in this frequency range  $D'' (\omega)$ is much larger than $D^{\prime}(\omega)$, and at the same time $|D^{\prime\prime} (\omega)| <1$  (see the small-$\omega$ behaviors in Fig.\ref{fig:realDelta}).  Keeping only
$D^{\prime\prime }$
in (\ref{eq:N}) and using $D^{\prime\prime} (-\omega) = - D^{\prime\prime} (\omega)$, we obtain
\begin{equation}
N(\omega) \approx \text{Re}\left(\frac{1}{\sqrt{1-|D^{\prime\prime}(\omega)|^{2}}}\right)
\label{eq:N1}
\end{equation}
  Obviously, $N(\omega)$ is enhanced in this frequency range, and the maximum enhancement is where $|D^{\prime\prime}(\omega)|$ is the largest.

  We also see  from Fig. \ref{fig:DOS_1} that as temperature increases, the peak position in $N(\omega)$ shifts to a smaller frequency. We verified that it vanishes at $T= T_{p}$. In ``high-Tc'' language, the DOS displays BCS-like  ``gap closing'' behavior.  This is in variance with the DOS  in the same $\gamma$-model for even frequency case. There, $N(\omega)$ displays the  ``gap filling'' behavior in a range of temperatures below $T_p$.  On a more careful look, we found that the difference is again related to the special role of fermions with $\omega = \pm \pi T$ for even frequency case and the lacking of such special role for odd-frequency case.

  We note in passing that in our case there is no peak in  $N(\omega)$ at $\omega =0$ (i.e., no zero bias anomaly).
        Such peak has been theoretically predicted for odd-frequency pairing in  heterostructures, like s-wave superconductor/ferromagnet or p-wave superconductor/normal metal.
        We argue therefore that  a  peak in $N(\omega)$ at $\omega =0$  can't be used as a generic fingerprint of odd-frequency  pairing.

\section{Multiple solutions of the gap equation}
\label{sub:multiple_solutions_gap_equation}

For even-frequency pairing, we have shown in recent publications~\cite{paper_1,paper_2,paper_3,paper_4,paper_5,paper_6}
  that the sigh-preserving  $\Delta(\omega_m)$,  is not the only solution of the gap equation at $T=0$, but rather the end point of an infinite discrete set of solutions. The other end point is the solution of the linearized gap equation, which still exists for $N < N_{cr}$, as we explicitly demonstrated in Ref. \cite{paper_1}.
 We now show that the same holds for
 odd-frequency pairing,
 i.e., that the sign-preserving
  $D(\omega_m)$ is the end point of a discrete but infinite set of solutions.  For a generic solution in the set, the function $D(\omega_m)$ changes sign $n$ times along the positive Matsubara axis.  We label such solution as $D_n (\omega_m)$. The sign-preserving solution is  then $D_0 (\omega_m)$.

  We present two elements of the proof that such set exists. First, we present the exact solution of the linearized gap
   equation for $N < N_{cr}$.  The solution  changes sign an infinite number of times along the positive Matsubara axis and in our nomenclature is $D_\infty (\omega_m)$. This is the other end point of the discrete set of $D_n (\omega_m)$.
    Second, we solve the linearized gap equation at a finite $T$ without restricting to sign-preserving solutions and find a number of onset temperatures $T_{p,n}$ for $D_n (\omega_m)$, which change sign $n$ times at positive $\omega_m$.
    Because each zero along the Matsubara axis is a center of a dynamical vortex, gap functions with different $n$ are topologically distinct, i.e., $D_n (\omega_m)$  cannot gradually transform into a gap function with some other $n$.

\subsection{Exact solution of the linearized gap equation at \texorpdfstring{$T=0$ and $N < N_{cr}$}{} }

Earlier we argued that at $N = N_{cr}$, the solution of the linearized gap equation at small frequencies:
 $D(\omega_m) \propto 1/|\omega_m|^{1-\gamma/2} (1 + c\log{|\omega_m|})$,  has a free parameter $c$. We demonstrated
  that this parameter can be chosen to obtain the solution at all frequencies, which smoothly interpolates between this form and the  high-frequency form $D(\omega_m) \sim 1/|\omega_m|^{2+\gamma}$.   This is an expected result as by general reasoning the linearized gap equation should  have a non-trivial solution right at $N = N_{cr}$.

  Let's now move to $N < N_{cr}$.  For superconductivity coming out of a Fermi liquid, it is natural to assume, even for  odd-frequency pairing, that there should be
  the solution of the non-linear gap equation, but
  no
   solution of the linearized gap equation.  However, in our case of pairing out of NFL, the situation is different.  As we found above, if we consider small frequencies (the ones for which $\Sigma (\omega_m) > \omega_m$), we find
   the solution of the linearized gap equation in the form
   \bea
   D_\infty(\omega_m) &=& \frac{1}{2 |\omega_m|^{1-\gamma/2}} \left(C |\omega_m|^{i\gamma \beta_N} +  C^* |\omega_m|^{-i\gamma \beta_N}\right) \nonumber \\
   && = \frac{|C|}{|\omega_m|^{1-\gamma/2}}\cos({\gamma \beta_N  \log{|\omega_m|} + \phi})
   \label{eq:sol_lin_a}
   \eea
   where $\beta_N$ is given by (\ref{eq:beta}) and $\phi$ is the phase of the infinitesimally small complex factor  $C = |C|e^{i\phi}$.
     We see that at  small frequencies $D_\infty(\omega_m)$ oscillates an infinite number of times as a function of $\log{\omega_m}$ (hence the subindex $n = \infty$).  But we also see that it  contains a free parameter - this time the phase $\phi$.
     The issue then is whether  one can choose a particular $\phi$ and obtain the solution for $D_\infty(\omega_m)$ at all frequencies,  like for $N = N_{cr}$.

  The answer is affirmative -- we found the exact solution of the linearized gap equation at $T=0$. The solution
   has the form of Eq. (\ref{eq:sol_lin_a}) at small frequencies, with some particular $\gamma-$dependent $\phi$, and scales as $D_\infty (\omega_m) \propto 1/|\omega_m|^{\gamma +2}$ at large frequencies.    The solution is obtained using the same computational procedure as for even-frequency pairing.
    We skip the details of the derivation (they follow Refs. \cite{paper_1,paper_4,paper_5,paper_6}) and  just present the result.

\begin{figure}
	\includegraphics[width=8.5cm]{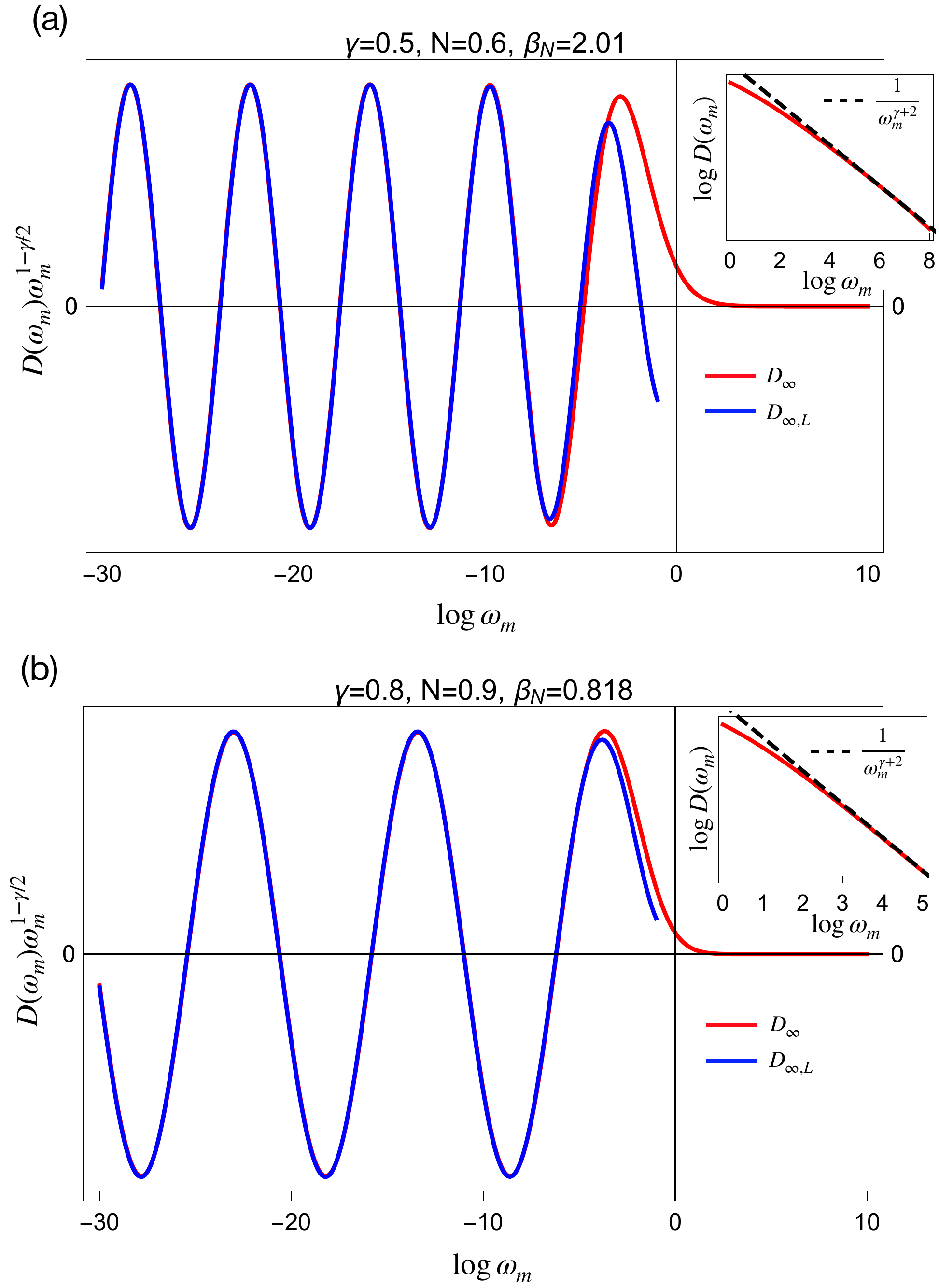}
	\caption{Red lines - the exact solution of the linearized gap equation, $D_\infty (\omega_m)$ in \eqref{eq:exact2},  at $T=0$  for  (a) $\gamma=0.5$,
$N=0.6 < N_{cr}=0.768$ and (b) $\gamma=0.8$ $N=0.9 < N_{cr}=0.945$. The insets show $D_\infty (\omega_m)$ at large frequencies.  The blue lines are approximate solutions $D_{\infty,L}(\omega_m)$, described in the text.  The approximate solution almost coincides with the exact one at small $\omega_m$, but becomes invalid for $\omega_m \geq 1$. }\label{fig:linearsolution}
\end{figure}
  The function $D_\infty (\omega_m)$ is expressed as
\begin{equation}
	D_{\infty}(\omega_m)=\frac{1}{|\omega_m|} \int_{-\infty}^\infty d\beta\frac{\cos[\beta\log(\omega_m^\gamma|1-\gamma|)+I(\beta)]}{\sqrt{\cosh[\pi(\beta-\beta_N)]\cosh[\pi(\beta+\beta_N)]}}\label{eq:exact2}
\end{equation}
where $\beta_N$ is the solution of \eqref{eq:beta} and
\begin{equation}
I(\beta)=\frac{1}{2}\int_{-\infty}^{\infty}d\beta'\log|1-\frac{1}{N}\epsilon_{\beta'}|\tanh(\pi(\beta'-\beta))
\label{eq:exact1}
\end{equation}
where $\epsilon_\beta$ is  given in \eqref{eq:beta}.   We plot $\Delta_\infty (\omega_m)$ in Fig.\ref{fig:linearsolution}.  We see that $\Delta_\infty (\omega_m)$ oscillates down to smallest frequencies, as a function of $\log |\omega_m|$ and decays as $1/|\omega_m|^{\gamma +2}$  at high frequencies.

To better understand the crossover between low-frequency and high-frequency behavior, we also plot in Fig.\ref{fig:linearsolution} the full `local' $D_{\infty, L} (\omega_m)$, which we obtained by adding to the low-frequency form(\ref{eq:sol_lin_a}) the series, obtained by expanding in $|\omega_m|^\gamma$, with the coefficients evaluated by restricting to contributions from internal $\omega_n$ comparable to $\omega_m$ (in practice this implies that we regularized formally UV divergent integrals by the $\Gamma$-functions,  see Ref.... for details). These series are
$D_{\infty, L} = |\omega|^{\gamma/2-1} {\text Re} \sum_{k=0}^\infty C_k |\omega_m|^{\gamma k}$, where $C_k = C \prod_{m=1}^k\frac{1}{I_\gamma(m)}$, and
\begin{equation}
	\begin{aligned}
		I_\gamma(m)=\frac{1}{2}&\left[\frac{1}{N}\left(B(1-\gamma,\frac{\gamma}{2}+\gamma m+i\gamma \beta_N)\right.\right.\\
		+&B(1-\gamma,\frac{\gamma}{2}-\gamma m-i\gamma \beta_N)\\
		-&\left.\left.B(\frac{\gamma}{2}+\gamma m+i\gamma \beta_N,\frac{\gamma}{2}-\gamma m-i\gamma \beta_N)\right)-\frac{2}{1-\gamma}\right]
	\end{aligned}\label{eq:Igamma}
\end{equation}
We see from  Fig.\ref{fig:linearsolution} that local series
 correctly describe the exact solution up to $\omega_m = O(1)$. The series, however, fail at higher  frequencies. In this range  the dominant contribution to $D_\infty  (\omega_m)$ comes from the non-local, UV terms, which  account for the crossover to high-frequency  behavior.

\subsection{Multiple onset temperatures \texorpdfstring{$T_{p,n}$ for $N < N_{cr}$}{} }

\begin{figure}
	\includegraphics[width=8.5cm]{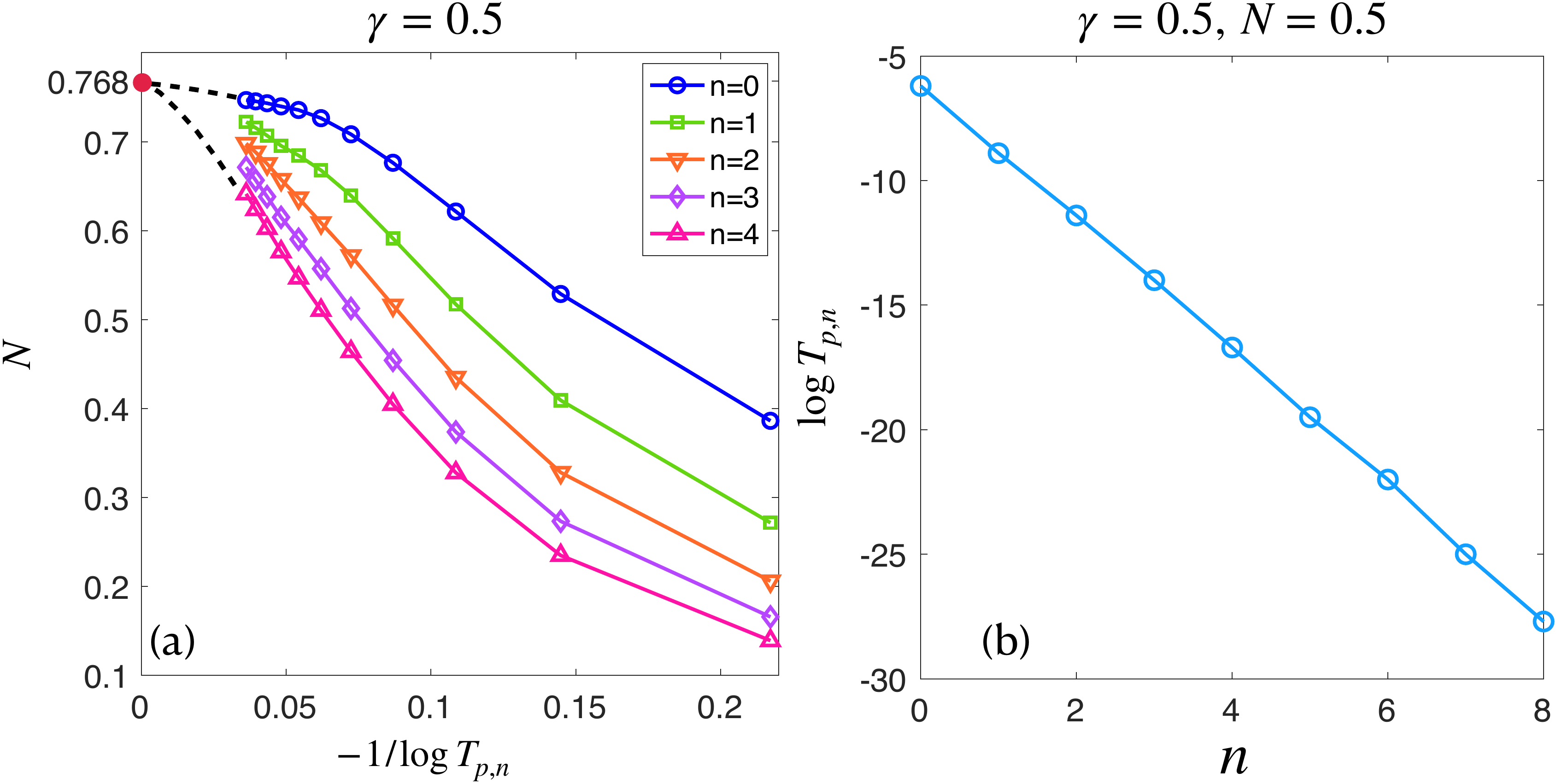}
	\caption{(a)The first five solutions to the linearized gap equation for $\gamma=0.5$ and at various temperatures. For better illustration we plot $N$ versus $-1/\log(T_{p,n})$. As $T\to0$ (and hence $-1/\log(T)\to0$), all curves will terminate at $N_{cr}=0.768$, consistent with $T=0$ analysis. (b) Linear dependence of $\log(T_{p,n})$ on $n$ for a particular $\gamma=0.5$ and $N=0.5$, which indicates $T_{p,n}$ decays with $n$ exponentially. }\label{fig:multiple-solutions}
\end{figure}
 The complimentary evidence for an infinite set of solutions comes from more sophisticated analysis of the linearized gap equation at a finite $T$.  We solved the matrix equation in Matsubara frequencies using the hybrid approach, which allows one to reach very low temperatures (see Ref\cite{paper2} for details).
  We found a number of onset temperatures,  $T_{p,n}$, in addition to $T_{p,0}$ in Fig.\ref{fig:Tp0}.
 The first few solutions for $n =0, 1,2,3$ and $4$ are shown in Fig.\ref{fig:multiple-solutions}(a).
  The corresponding eigenfunction $\Delta_n (\omega_m)$ changes sign $n$ times along the positive Matsubara axis,  which leaves little doubt that  $T_{p,n}$ is the onset temperature for the $n$-th member of the infinite set at $T=0$.
Although difficult to obtain zero temperature result, from
Fig.\ref{fig:multiple-solutions}(a) we can see that all $T_{p,n}$ smoothly extrapolates to $N = N_{cr}$ at $T=0$.  This implies that $N = N_{cr}$, $T=0$ is a critical point of an infinite-order, below which an infinite number of solutions emerges simultaneously. Moreover, Fig.\ref{fig:multiple-solutions}(b) shows that for a particular $\gamma$ and $N$, $T_{p,n}$ decays exponentially with increasing $n$.

\section{Case \texorpdfstring{$\gamma>1$}{} }
\label{sec:the_case_}

  We see from Fig.\ref{fig:Ncr} that $N_{cr}$ approaches 1 at $\gamma \to 1$.
 The issue we discuss now is what happens for larger values of the exponent $\gamma$.
  A formal answer is that the line $N_{cr} (\gamma)$ coincides with $N=1$ for all $\gamma >1$,
   and the transition upon variation of $N$ becomes strongly first order: for $N >1$, the only solution of the gap equation (\ref{eq:Delta_N1T0})  is $D(\omega_m) =0$, while for $N <1$ it is $D(\omega_m) = \infty $  for any $\omega_m$.  This bizarre behavior comes about because the gap equation at $T=0$ and $N \neq 1$ contains a divergent piece.

On a more careful look we note that the divergence in the gap equation at $T=0$ is an unphysical element of Eliashberg theory at a QCP as it  originates from the divergence in the fermionic self-energy. The latter must be regularized to avoid an unphysical behavior of the spectral function. A way to do this is to analyze Eliashberg theory either at a finite $T$ or a finite bosonic mass $\omega_D$ and treat $T=0$ theory at a QCP as a limit when $T$ and/or $\omega_D$ are vanishingly small but still finite. As long as $T$ or $\omega_D$ are non-zero, both the self-energy and the gap equation are free from divergencies.

 Below we apply this strategy to analyze the gap equation at a generic $N$. We show in Sec.~\ref{sec:sol1} that
 there is a set of continuous transitions at $1/N_{cr,n}-1 \propto T^{\gamma-1}$ when $T$ is used as a regularization, or  $1/N_{cr,n}-1 \propto \omega_D^{\gamma-1}$, when $\omega_D$ is used as a regularization. Once $N$ gets smaller than $N_{cr,n}$, the system becomes unstable towards  pairing with a gap function $D_n (\omega_m)$ with $n$ nodes along the positive Matsubara half-axis.

We next show that on top of this, there are new features in the system behavior at $\gamma >2$.  Namely, pairing correlations become nearly divergent already at $N < N_{cr,n}$, and the pairing  gap remains very small up to some other $N > N_{cr,n}$ and then rapidly increases. In other words, at vanishing $T$ and $\omega_D$, the system remains frozen at the critical  point towards  pairing with $D_n (\omega_m)$ in some range of $N$ around $N_{cr, n}$.  To see this clearly, in  Sec.~\ref{sec:sol2} we introduce an additional parameter into the $\gamma$ model and obtain a generalized phase diagram,
in which the emergence of the range
around
$N_{cr,n}$ can be seen as a flattening of particular critical line.

\subsection{Regularization by a finite temperature and/or bosonic mass}
\label{subsubsec:M_b0_finiteT}

Like we said, the r.h.s. of the  gap equation~(\ref{eq:Delta_N1T0})  diverges at $N \neq 1$. The divergence comes  from the integration over $\omega_n \approx \omega_m$. Approximating the numerator in Eq.~(\ref{eq:Delta_N1T0}) by its value at $\omega_n = \omega_m$ and pulling it out of the integral over $\omega_m$, we obtain the singular piece in the r.h.s of  Eq.~(\ref{eq:Delta_N1T0}) in the form
\beq
\frac{1-N}{2N} \frac{D (\omega_m) \text{sign} \omega_m}{\sqrt{1 + D^2 (\omega_m)}} \int \frac{d\omega_n}{|\omega_m - \omega_n|^\gamma}.
\label{eq:a}
\eeq
There are two ways to regularize this divergence: one can either keep $T$ small but finite, or keep a non-zero bosonic mass (i.e., move the system  slightly away from a QCP).  In the first case, $\int d \omega_m$ is replaced by $2\pi T \sum_{n \neq m}$ without the $n=m$ term as the latter cancels  out (see Sec.\ref{sec:model}).
In the second case, the gap equation at $T=0$ is obtained by replacing  $1/{|\omega_m - \omega_n|^\gamma}$ with  $1/[(\omega_m - \omega_n)^2 + \omega_D^2]^{\gamma/2}$, in which case,  the integral in (\ref{eq:a}) becomes  of order $1/(\omega_D)^{\gamma -1}$. When $T$ and $\omega_D$ are both finite, the gap equation becomes
\begin{equation}
	D(\omega_m)\omega_m= \pi T \sum_{\omega_n} \frac{\frac{1}{N}D(\omega_n)-D(\omega_m)}{\sqrt{1+D^2(\omega_n)}} \frac{\sgn(\omega_n)}{|(\omega_m-\omega_n)^2+\omega_D^2|^{\gamma/2}}.\label{eq:TMb}
\end{equation}
For small $T$, $\omega_D$ and $1-N$, we keep these terms only in the would-be-divergent piece. Moving it
 to the l.h.s., one can approximate the gap equation  as
\begin{align} \label{eq:delta3}
     & D(\omega_m) \left( \omega_m - \frac{a_\gamma}{\sqrt{1 + D^2 (\omega_m)}} \right)=
      \frac{1}{2} \int d \omega_m \frac{D(\omega_n)-D(\omega_m)}{\sqrt{1 + D^2 (\omega_n)}} \frac{\sgn(\omega_n)}{
      |\omega_m-\omega_n|^\gamma},
   \end{align}
where
\begin{equation}
 \begin{aligned}
    a_\gamma &=\frac{1/N-1}{(2\pi T)^{\gamma-1}} \zeta (\gamma) ~&\text{for regularization by } T  \\
    a_\gamma &=\frac{1/N-1}{\omega_D^{\gamma-1}}\frac{\sqrt{\pi}\Gamma[(\gamma-1)/2]}{2\Gamma(\gamma/2)}~&\text{for regularization by } \omega_D\\
 \end{aligned}\label{eq:reg}
\end{equation}
We verified numerically that  Eqs.~(\ref{eq:TMb}) and (\ref{eq:delta3}) have identical solutions at $a_\gamma = O(1)$, which are relevant to our discussion.  Below we chiefly present the solution of Eq.~(\ref{eq:delta3}).

\subsection{Solution of the regularized gap equation} \label{sec:sol1}

To obtain the boundary of the region where $D(\omega_m)$ is non-zero, we
   again analyze
  the linearized gap equation, i.e., Eq. (\ref{eq:delta3}) with infinitesimally small $D(\omega_m)$.
     Like before, we  search for power-law solution
  $D(\omega_m) \propto |\omega_m|^\alpha$ at small frequencies. We obtain
   $\alpha_1=0$ and $\alpha_2 = \gamma-2$, independent on $a_\gamma$.  We then note that
   a non-zero $T$ and/or $\omega_D$ sets the boundary condition that  $D(\omega_m)$ must be a constant at $\omega_m =0$ and its expansion around $\omega_m =0$ must be analytic.  This selects the single solution with $\alpha_1 =0$.

    For $\gamma < 1$ at $N > N_{cr}$ we also selected a single solution with a real exponent at small $\omega_m$, and then
      argued that this solution  cannot be smoothly connected to another power-law behavior at high frequencies, hence
      the solution of the linearized gap equation does not exist.
       Here, the situation is less obvious because in Eq. (\ref{eq:delta3}) we have the parameter $a_\gamma$ which we can vary.  This $a_\gamma$ does not affect the behavior of $D(\omega_m)$ at both small and large frequencies, but  controls the transformation between small and high-frequency forms.

    \begin{figure*}
\includegraphics[width=15cm]{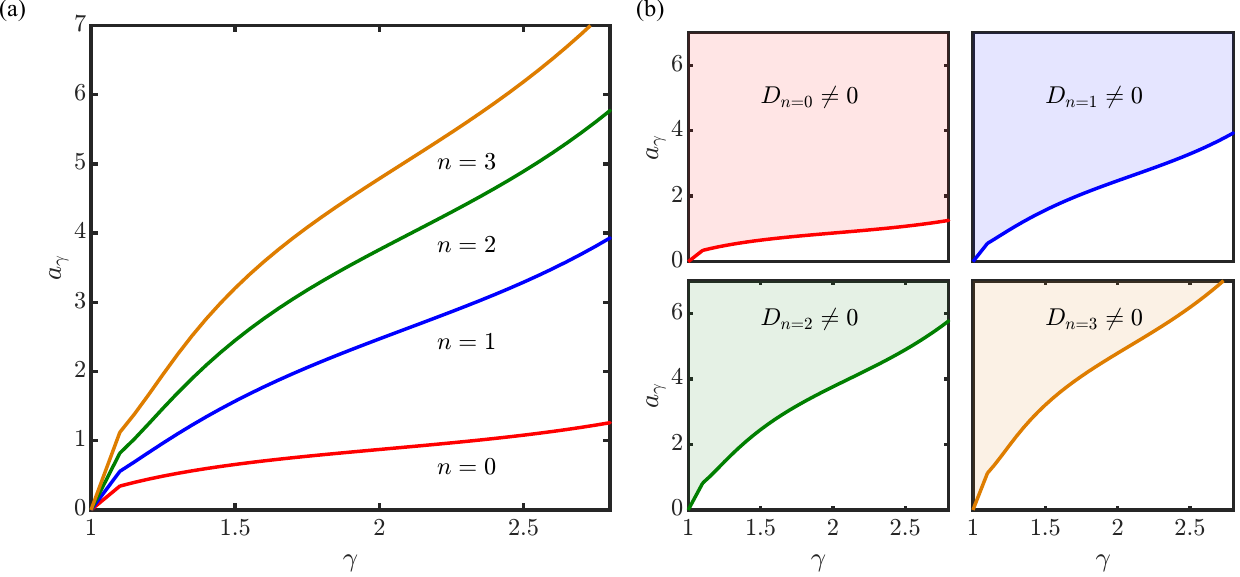}
\caption{Left panel: the result of numerical analysis of the gap equation  (\ref{eq:delta3}) for infinitesimally small $D(\omega_m)$ on the $(a_\gamma, \gamma)$ plane.  The solution exists at a set of lines $a_{\gamma,n}$.
 Above each line, there appears a non-zero $D_n (\omega_m)$ with $n$ nodes at positive $\omega_n$. We illustrate this in the right panel.}\label{fig:pd_0}
\end{figure*}

  \begin{figure}
  \includegraphics[width=8cm]{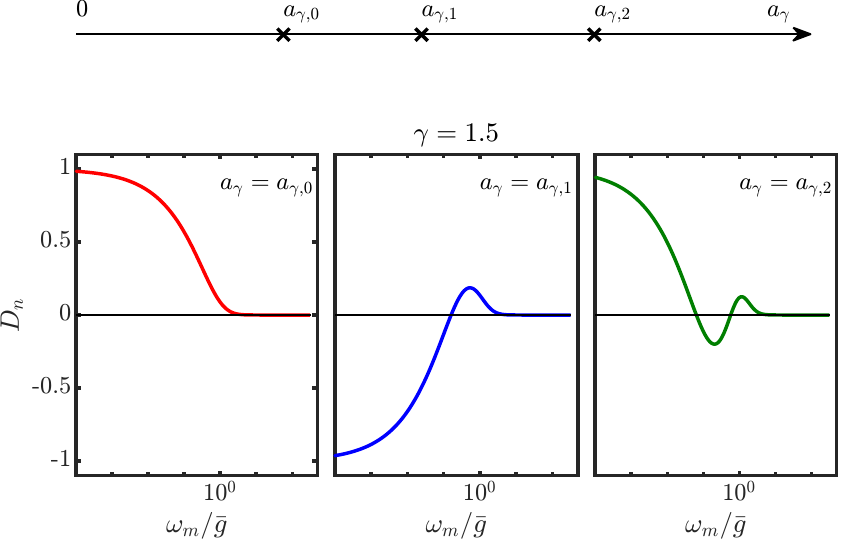}
  \caption{The gap functions $D_n (\omega_m)$ for critical $a_{\gamma,n}$ with $n=0,1,2$ for representative $\gamma =1.5$.
  }
  \label{fig:fig1}
  \end{figure}

    \begin{figure*}
\includegraphics[width=15cm]{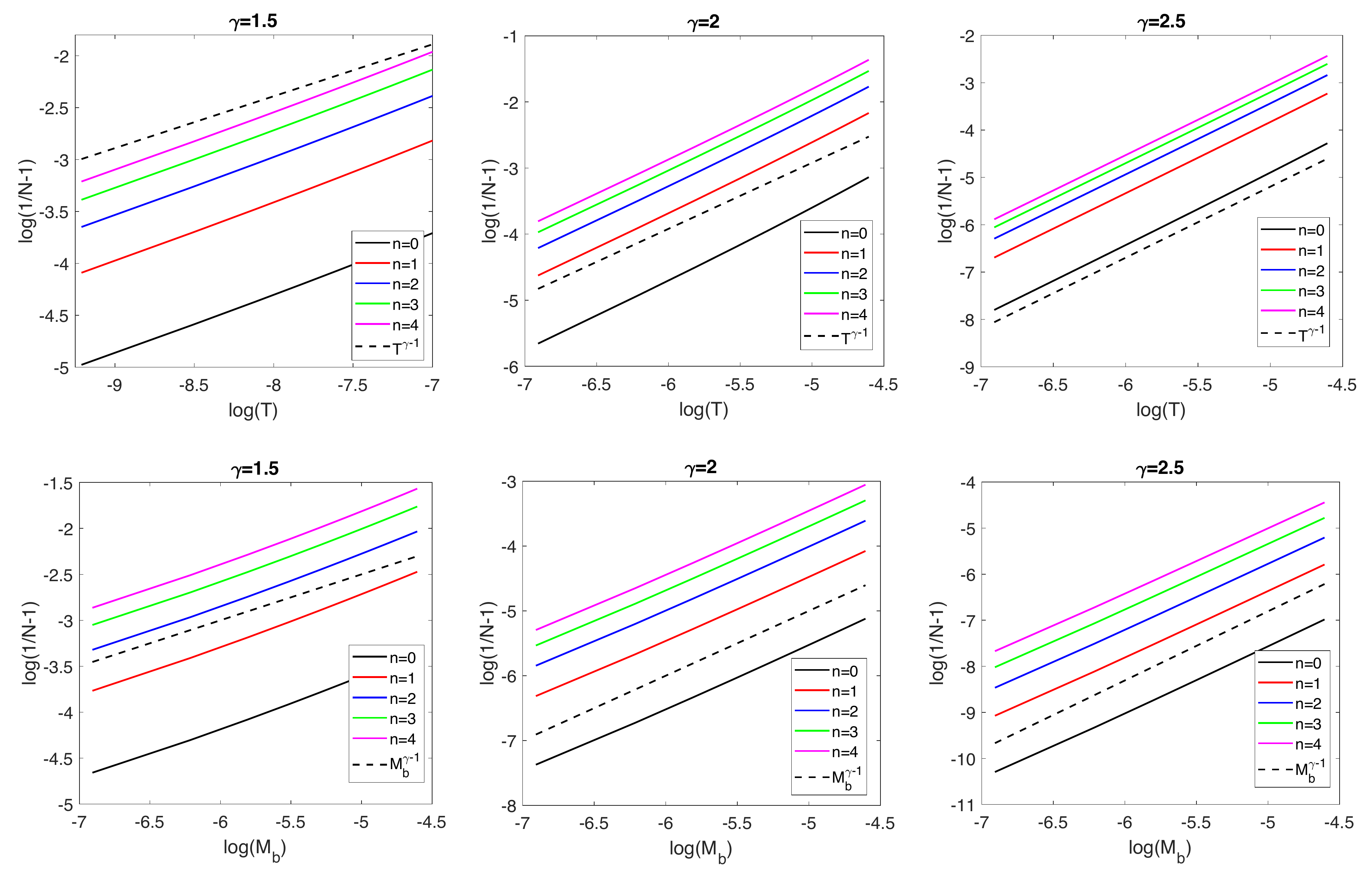}
\caption{The location of the onset lines for the first five solutions  in $(N, T)$ and $(N, \omega_D)$ planes (upper and lower panels) for three representative values of $\gamma >1$, obtained by solving numerically the gap equation
 (\ref{eq:TMb}).
  We set $\omega_D=0$ in the upper panel and  set $T=5\times10^{-5}$ in the lower panel. We see that the
   critical lines follow $T_{p,n} \propto (1/N-1)^{1/(\gamma-1)}$
   and $\omega_{D,n}  \propto (1/N-1)^{1/(\gamma-1)}$, as expected from the solution of the simplified Eq. (\ref{eq:delta3}) (see Fig. \ref{fig:pd_0}).
     The plots are in log-log scale, the dashed lines are guides to the scaling behavior}\label{fig:TpN}
\end{figure*}

In Fig.\ref{fig:pd_0}, we present
the numerical solution of the linearized gap equation. We see that there is a discrete set of lines in the $(a_\gamma, \gamma)$ plane, where the solution satisfying boundary conditions exists. Above each line, there appear a non-zero $D_n (\omega_m)$ with $n$ nodes at positive $\omega_n$.  We demonstrate this in Fig.~\ref{fig:fig1} for a representative $\gamma=1.5$.
These gap functions are topologically distinct, which implies that they can be treated separately.
 Note that the critical lines $a_{\gamma,n}$ all emerge from $a_\gamma =0$ at $\gamma=1$ and evolve continuously  with $\gamma$.
 In Fig. ~\ref{fig:TpN} we present the phase boundaries by solving the gap equation~(\ref{eq:TMb}) in the original variables $T, \omega_D$ and $N$. We see that the
   critical lines follow $T_{p,n} \propto (1/N-1)^{1/(\gamma-1)}$ and $\omega_{D,n}  \propto (1/N-1)^{1/(\gamma-1)}$, as expected from Fig. \ref{fig:pd_0}. The canonical model with $N=1$  remains in the normal state, but the pairing instabilities develop for any $N <1$  at proper $T$ and/or $\omega_D$.

\begin{figure}
  \includegraphics[width=8cm]{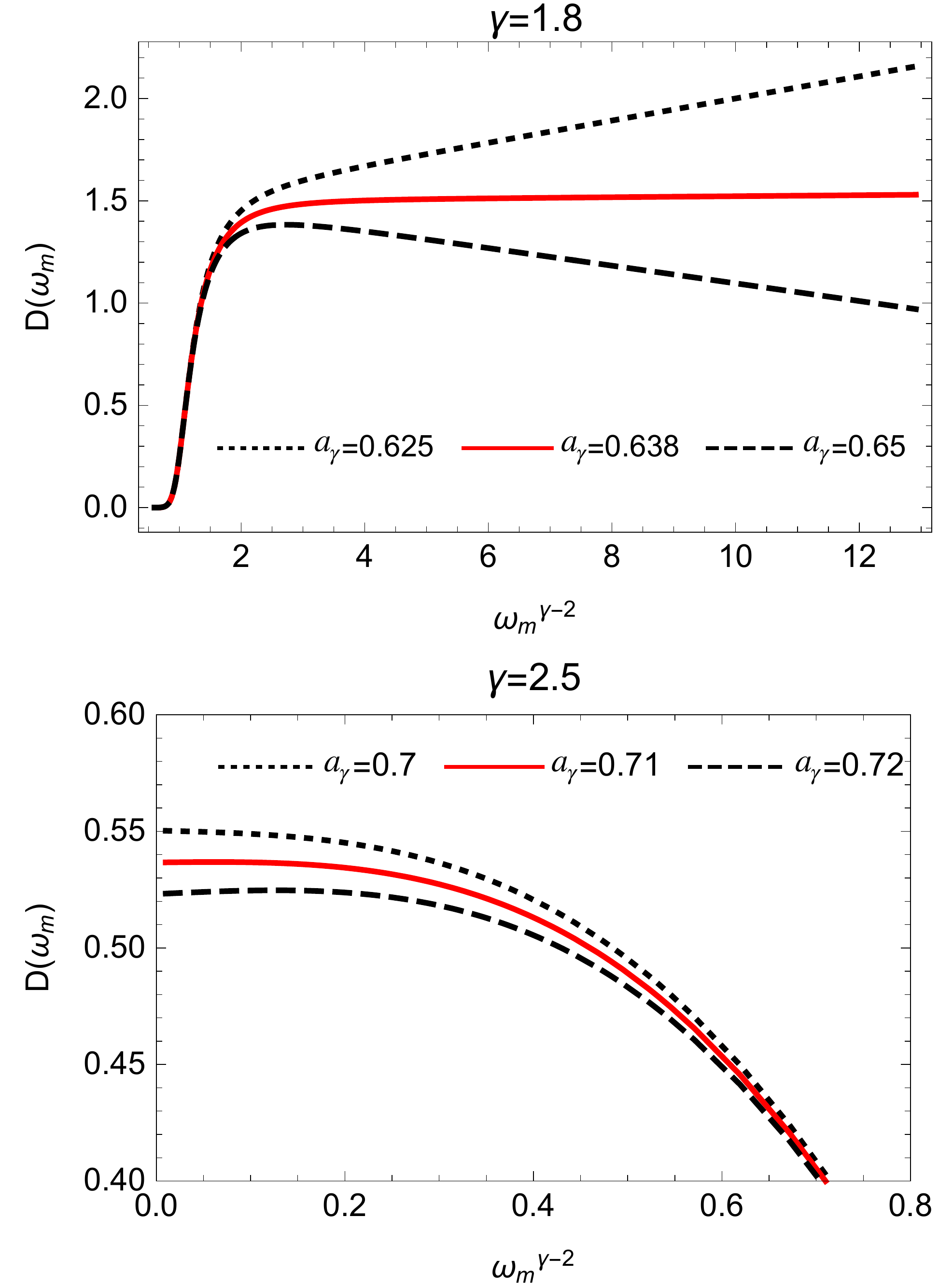}
  \caption{The first solution $D_0 (\omega_m)$ of Eq.\eqref{eq:III} for $\gamma =1.8$ and $\gamma =2.5$. The red curves show the correct solutions which satisfy the boundary condition. This happens when $a_\gamma=a_{\gamma,0}$ so we have $a_{\gamma,0}=0.638$ for $\gamma=1.8$ and $a_{\gamma,0}=0.71$ for $\gamma=2.5$. For comparison, we also show the eigenfunctions with $a_\gamma$ slightly deviating from these critical values. }\label{fig:diffeqs}
\end{figure}

\begin{figure}
  \includegraphics[width=8cm]{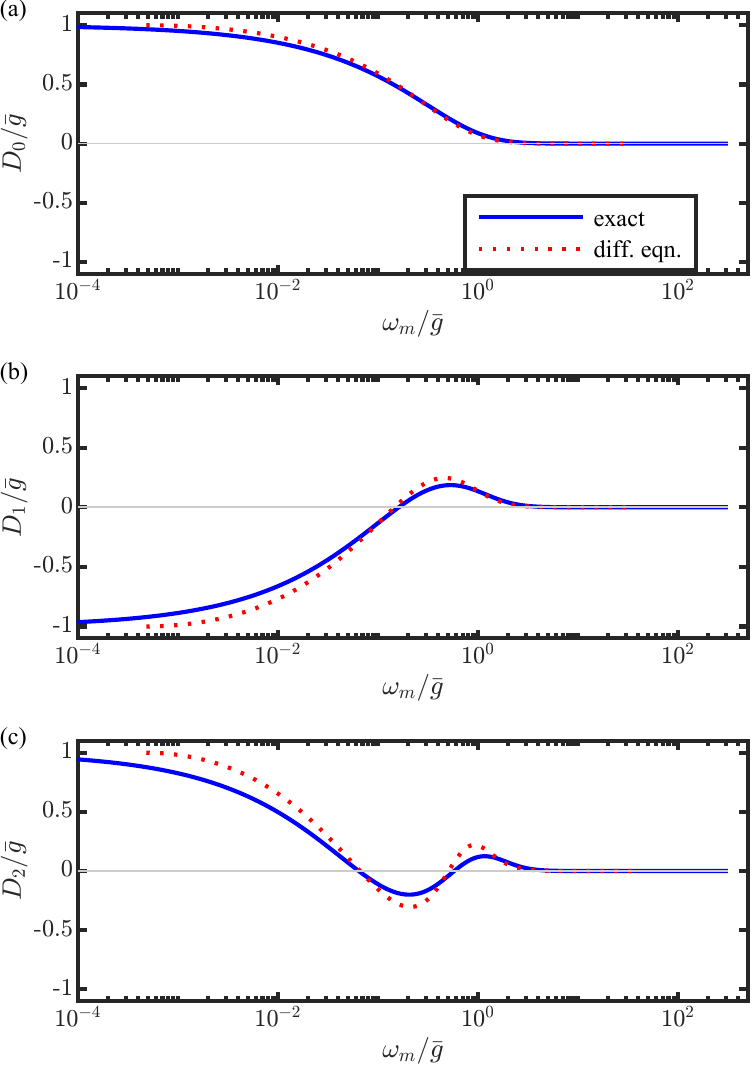}
  \caption{The gap functions $D_n(\omega_m)$ obtained by solving the differential equation Eq.~(\ref{eq:III}) (red dotted line) and the original gap equation Eq.~(\ref{eq:delta3}) (blue solid line). The critical value of $a_{\gamma,n}$ are $0.557836, 1.31383, 1.96301$ in the former case and $0.65826, 1.5775, 2.4586$ in the latter case. }
  \label{fig:diffeqs2}
\end{figure}

  The existence of a discrete set of $a_{\gamma,n}$ can be understood analytically, at least at a qualitative level.
   For this we approximate $\int d\omega_n\sgn\omega_n/|\omega_m-\omega_n|^\gamma\approx2\gamma(\int_0^{\omega_m} d\omega_n\omega_n/\omega_m^{\gamma+1}+\int_{\omega_m}^\infty d\omega_n\omega_m/\omega_n^{\gamma+1})$
   and convert the  integral gap equation (\ref{eq:delta3}) into the differential one.  The linearized differential  gap equation is
   \begin{equation}
	\begin{aligned}
		&\bar{D}''(x)x^2(x+1+\gamma/2-a_\gamma x^{1-1/\gamma})\\
		+&\bar{D}'(x)\frac{2x}{\gamma}\left((\gamma-1)x-1-\gamma/2-a_\gamma(\gamma-2)x^{1-1/\gamma}\right)\\
		-&\bar{D}(x)\frac{2}{\gamma}\left(x-1-\gamma/2-\frac{3(\gamma-1)}{2\gamma}a_\gamma x^{1-1/\gamma}\right)=0
	\end{aligned}\label{eq:III}
\end{equation}
where $x=|\omega_m|^\gamma$ and  $\bar{D}(x)=D(x) x^{2/\gamma}$. The two boundary conditions are
$\bar{D}(x) \propto x^{2/\gamma}$  at small $x$ (i.e., $D(\omega_m) = {\text const}$) and $\bar{D}(x) \propto 1/x$
(i.e., $D(\omega_m) \propto 1/\rvert\omega_m\rvert^{2 +\gamma}$) at large $x$.
A generic solution of Eq.~(\ref{eq:III}) can be readily obtained numerically for arbitrary $a_\gamma$ and even
analytically at $a_{\gamma}=0$.   For a generic $a_\gamma$, a solution that satisfies the boundary condition at small $x$ does not satisfy the one at large $x$. However, for a discrete set of $a_\gamma = a_{\gamma,n}$, we did find the solutions that satisfy both boundary conditions. The corresponding $\bar{D}(x)$ change sign $n$ times at positive $x$, in full agreement with the earlier analysis
\footnote{The term with the first derivative can be eliminated by changing variables, after which
  the differential equation in Eq. \eqref{eq:III} becomes an effective Schr{\"o}dinger equation $-\psi ''(x)+V_{a_{\gamma }}(x)\psi (x)=0$ along the half-axis $x \geq 0$.
Here $V_{a_{\gamma }}(x) = [g_2(x)/g_1(x)]^2/4 - g_3(x)/g_1(x) + [g_2(x)/g_1(x)]^{\prime}/2$ is the
 effective  potential, which depends on
 $a_{\gamma }$ and in the two limits reduce to
$V(x=0)= + \infty$
and $V(x\rightarrow +\infty )\rightarrow 0$. The functions $g_{i}(x)$  ($i=1,2,3$) are the  terms multiplying ${\bar D}^{\prime\prime}(x)$, ${\bar D}^{\prime}(x)$ and ${\bar D}(x)$ in Eq.~(\ref{eq:III}), respectively.
 Like in the Schr{\"o}dinger equation, the discrete set of $a_{\gamma,n}$ is selected by the requirement that  $\bar{D}(x) \propto 1/x$ at large $x$, i.e., the solution is normalizable.}
As an illustration, in Fig.\ref{fig:diffeqs} we show the gap function $D_0 (\omega_m)$ for two values of $\gamma$ at $a_\gamma \approx a_{\gamma,0}$.  We clearly see that there exist particular  $a_\gamma \equiv
a_{\gamma,0}$, when the gap function satisfies the boundary conditions at $x =0$ and $x \to \infty$.
In Fig.~\ref{fig:diffeqs2}, we compare the solutions of the differential and integral gap equations for $n=0,1$ and $2$.
with the one from solving the original gap equation Eq.~(\ref{eq:delta3}). We see that the corresponding gap functions agree, except for irrelevant fine features.

\subsection{Hidden degeneracy at \texorpdfstring{$\gamma >2$}{}.}\label{sec:sol2}

Fig.~\ref{fig:pd_0} shows that critical lines $a_{\gamma,n}$ gradually pass  through $\gamma =2$. At a first glance this seems quite
 natural as the small-frequency form $D(\omega_m) = {\text const}$, which we used as a boundary condition, imposed by finite $T$ and/or $\omega_D$,  holds for  all  values of $\gamma >1$. At the same time, there is a special behavior
 at $\gamma =2$: the exponents $\alpha_1=0$ and $\alpha_2 =\gamma-2$ merge, and the two low-frequency forms become
  $D (\omega_m) = {\text const}$ and $D(\omega_m) = \log({|\omega_m|}/{\bar g})$.  The merging of the two exponents and the emergence of a $\log({|\omega_m|}/{\bar g})$  is similar to what we earlier found  at the critical $N$ at $\gamma <1$ and identified with the onset of an order.  There is a qualitative difference between that case of $N = N_{cr}$, $\gamma <1$   and the present case $N=1$, $\gamma =2$ in that at $N < N_{cr}$ the two exponents became complex, while here the exponents remain real for $\gamma >2$ and just pass through each other.

  \begin{figure}
  \includegraphics[width=8cm]{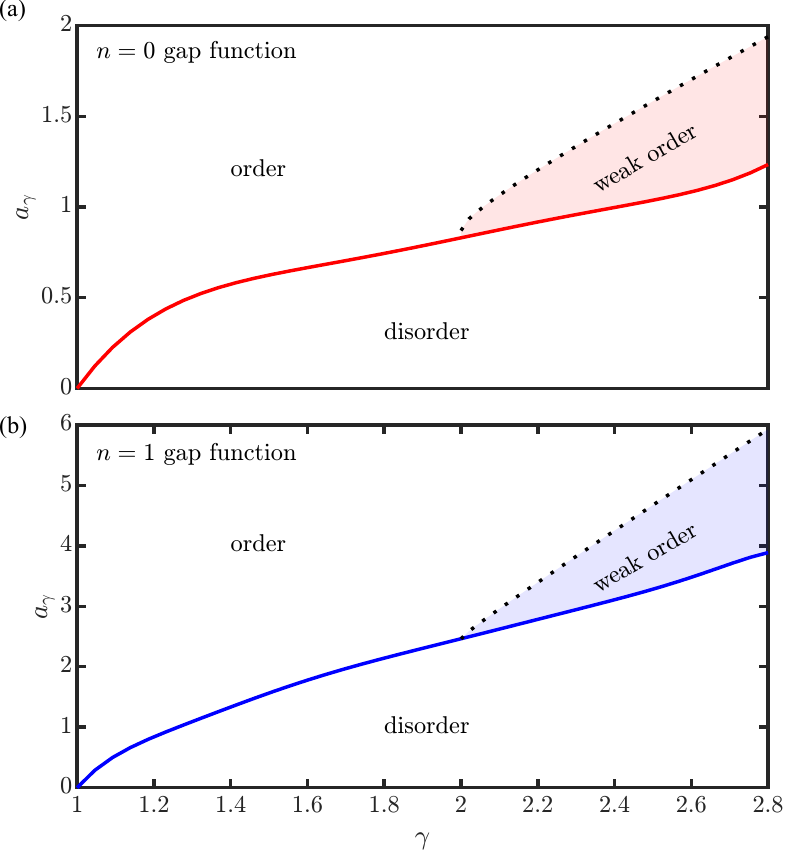}
  \caption{The modified phase diagram in $(a_\gamma, \gamma)$ plane.   A non-zero $D_n (\omega_m)$ still emerges
   above $a_{\gamma,n}$, as in Fig. \ref{fig:pd_0}, but the magnitude of $D_n (\omega_m)$  remains small in the shaded region and vanishes at $T=\omega_D \to 0$. Outside the shaded region, $D_n (\omega_m)$ remains finite at $T=\omega_D  \to 0$. }
  \label{fig:pd_0B}
\end{figure}

  Still, the merging of the exponents $\alpha_1$ and $\alpha_2$  at $\gamma =2$ suggests to look at the  system behavior around $\gamma =2$ more carefully.  Below we argue that while  the phase diagram in Fig. \ref{fig:pd_0} with lines
   $a_{\gamma,n}$, separating ordered and disordered states within each topological sector, is correct at $T >0$ or $\omega_D >0$ for any value of $\gamma$, the behavior near each of these lines is different between $\gamma <2$ and $\gamma >2$.
    Namely, for $\gamma >2$, there exists a finite range of $a_\gamma$ above each line, where $D_n (\omega_m)$ is finite but its magnitude is small in $T$ and/or $\omega_D$ and vanishes when $T, \omega_D \to 0$. We show this in Fig.~\ref{fig:pd_0B}.

    To understand this behavior, we further extend the model by additionally splitting  interactions in the particle-particle and particle-hole channel in such a way that this does not introduce divergencies for $\gamma >1$.
     This has been introduced in Refs.~\cite{paper_3,paper_5,paper_6} as a divergencies-free extension to $M \neq 1$ of the $\gamma$-model for $N=1$ and $T=\omega_D =0$.  Here, we apply the same extension to the model with a finite, but small $N \neq 1$ and small $T$ and/or $\omega_D$, i.e., to a model with  finite $a_\gamma$.    Our goal is to introduce a parameter which would allows us to vary the linearized gap equation near $a_{\gamma,n}$ and detect properties, which are not visible in the original model.

   The extension to $M \neq 1$ and the  derivation of the gap equation for the
   original
   model with $N=1$  is presented in the Appendix C in Ref.~\cite{paper_3}.  Performing the same analysis for the model  with a non-zero $a_\gamma$,
    we extend the gap equation (\ref{eq:delta3}) to
   \bea
&&D (\omega_m) \left(\omega_m - a_{\gamma} + \frac{1-M}{2} \int ~\frac{d \omega'_{m}}{|\omega_m - \omega'_{m}|^{\gamma}}  \left(\frac{\text{sgn} \omega_{m}}{\sqrt{1 + D^2 (\omega_{m})}} - \frac{\text{sgn} \omega'_{m}}{\sqrt{1 + D^2 (\omega'_{m})}}
\right)\right) = \nonumber \\
&& \frac{1}{2} \int ~\frac{d \omega'_{m}}{|\omega_m - \omega'_{m}|^{\gamma}}  \frac{D(\omega'_{m})-D(\omega_m)}{\sqrt{1 + D^2 (\omega'_{m})}} \text{sgn} \omega'_{m},
\label{3_12b}
\eea
where the frequency unit has been taken as ${\bar g}/M^{1/\gamma}$.
For  $M=1$, Eq. (\ref{3_12b}) reduces to (\ref{eq:delta3}). Note that this gap equation is free from infra-red singularities at any $M$.

Our goal is to obtain the phase diagrams in $(M, a_\gamma)$ plane for $\gamma <2$, $\gamma =2$, and $\gamma >2$, analyze
 the behavior at $M$ close to $1$, and extract from that additional features in the phase diagram of the non-extended model with $M=1$.

 \begin{figure}
  \includegraphics[width=8cm]{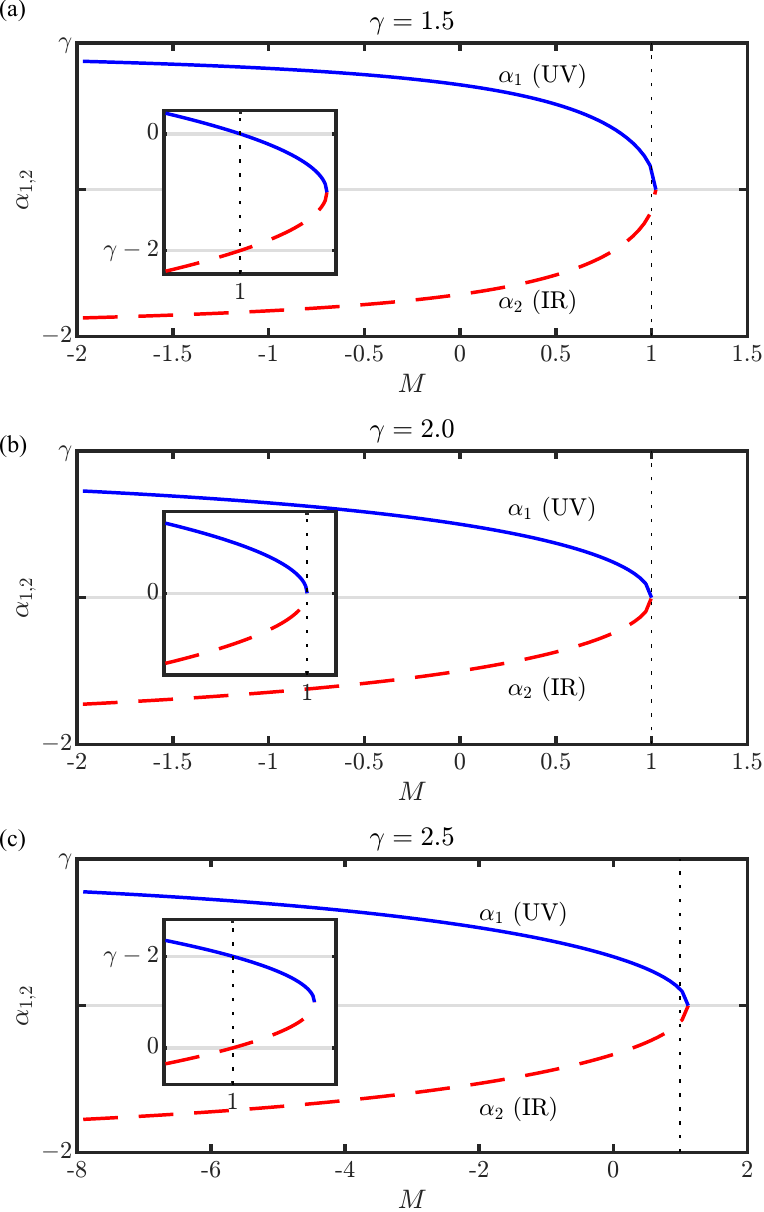}
  \caption{The exponent $\alpha_1$ (blue solid) and $\alpha_2$ (red dashed) of the possible power-law solutions at low-frequency of the gap equation as a function of $M$, where (a) $\gamma=1.5$, (b) $\gamma=2$, and (c) $\gamma=2.5$. Insets are zoom in around $M=1$.}\label{fig:exp}
\end{figure}

\begin{figure}
  \includegraphics[width=8cm]{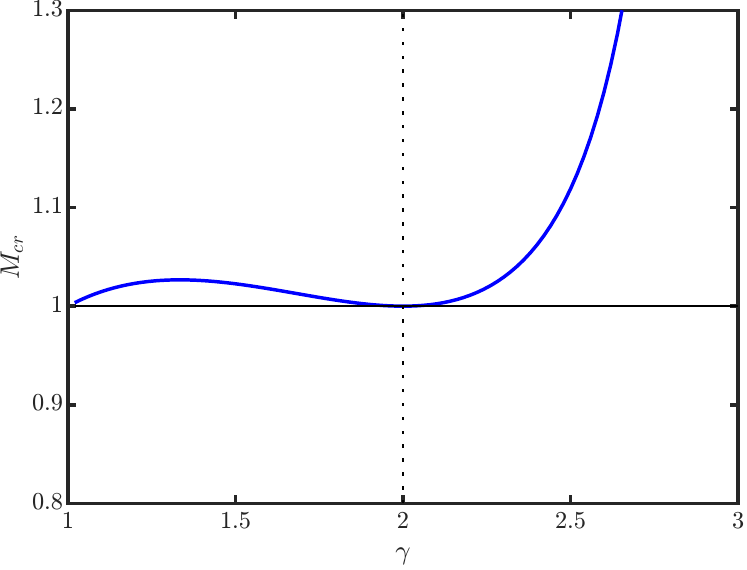}
  \caption{The boundary line $M_{cr} (\gamma)$ on the $(M, \gamma)$ plane.  The ordered state exists for $M > M_{cr}$.
      Pairing in a topological sector, specified by $n$, actually develops earlier $a_\gamma$  exceeds some threshold value (see text).}\label{fig:Mcr}
\end{figure}

 Like we did earlier, we consider the linearized gap equation at small frequencies.   One can easily verify that
  the two exponents $\alpha_1$ and $\alpha_2$ do depend on $M$ and are the solutions of
  \beq
M=\frac{\Gamma (2-\gamma ) \Gamma (-\alpha +\gamma -1)}{\Gamma (-\alpha )}+\frac{\pi  \Gamma (\alpha +1) (\csc (\pi  (\alpha -\gamma ))-\csc (\pi  \gamma ))}{\Gamma (\gamma -1) \Gamma (\alpha -\gamma +2)}-1  .
\label{s_1}
  \eeq
  We show the exponents $\alpha_1$ and $\alpha_2$ as functions of $M$ in Fig.~\ref{fig:exp}.
    Following the exponents all the way to $M \to - \infty$, where the pairing interaction is weak and a non-zero $\alpha$ emerges due to either ultra-violet (UV) singularity (internal $\omega'_m$ are much larger than external $\omega_m$) or IR singularity (internal $\omega'_m$ are much smaller than external $\omega_m$), we see that $\alpha_1 \approx \gamma$ is an UV exponent and $\alpha_2 \approx -2$ is an IR exponent.
  The two exponents remain real as long as $M < M_{cr}$,
  where
\beq
M_{cr}=\frac{\pi  \left(\csc \left(\pi  \gamma /2\right)-2 \csc (\pi  \gamma )\right) \Gamma \left(\gamma /2\right)}{\Gamma \left(1-\gamma /2\right) \Gamma (\gamma -1)}-1.
  \label{s_1_1}
  \eeq
  At a critical $M$,  the two exponents merge at  $\alpha_{1,2} =\gamma /2 -1$ and become complex at $M > M_{cr}$, which is the same behavior as around $N=N_{cr}$ for $\gamma <1$.  We plot $M_{cr}$ as a function of $\gamma$ in Fig.~\ref{fig:Mcr}. We see that $M_{cr}$ is larger than $1$ for both  $\gamma<2$ and $\gamma >2$, but is equal to $1$ at $\gamma =2$.
 In analytical form, the second derivative of the function $M(\alpha ,\gamma )$ over $\alpha $ at $\alpha = \gamma/2-1$ is
$$
\left.\frac{\partial^{2}M}{\partial \alpha^{2}} \right|_{\alpha =-1+\gamma /2}=
\frac{2 \pi  \sin ^2\left(\frac{\pi  \gamma }{4}\right) \Gamma \left(\frac{\gamma }{2}\right) \left(\pi ^2 \csc ^3\left(\frac{\pi  \gamma }{2}\right)+2 \csc (\pi  \gamma ) \left(\psi ^{(1)}\left(1-\frac{\gamma }{2}\right)-\psi ^{(1)}\left(\frac{\gamma }{2}\right)\right)\right)}{\Gamma \left(1-\frac{\gamma }{2}\right) \Gamma (\gamma -1)}
$$
It is zero at $\gamma =1$, monotonically decreases for $\gamma >1$ and diverges as $\frac{4}{\gamma -3}$ at $\gamma =3$.

  \begin{figure}
  \includegraphics[width=\textwidth]{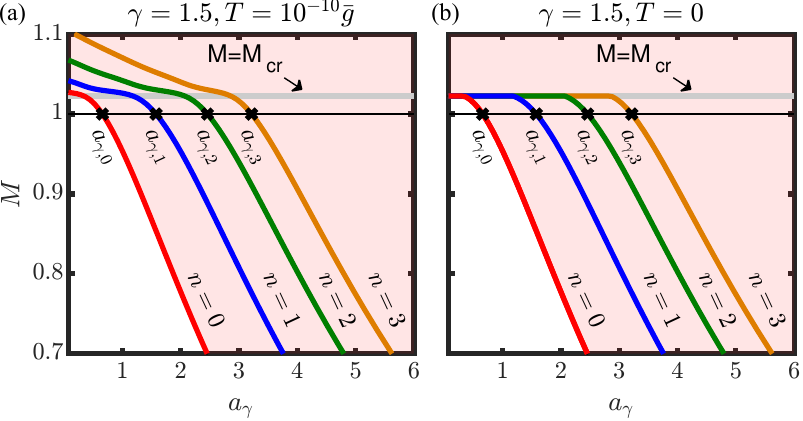}
  \caption{The phase diagram in the $(M, a_\gamma)$ plane for $\gamma =1.5$. There is a set of critical  lines $a_{\gamma,n} (M)$, above which pairing develops in a topological sector specified by $n$.
   (a) Numerical solution of Eq. (\ref{3_12b}) at $T =10^{-10}\bar{g}$,
   $\omega_D=0$; (b) expected result at $T = 0+$, $\omega_D=0$. In the last case we expect all $a_{\gamma,n} (M)$ to reach $M = M_{cr}$ at some minimal $n-$dependent $a_\gamma$ and remain at $M_{cr}$ at smaller $a_\gamma$.}
  \label{fig:pd_lt2_v2}
\end{figure}

  \begin{figure}
  \includegraphics[width=0.9\textwidth]{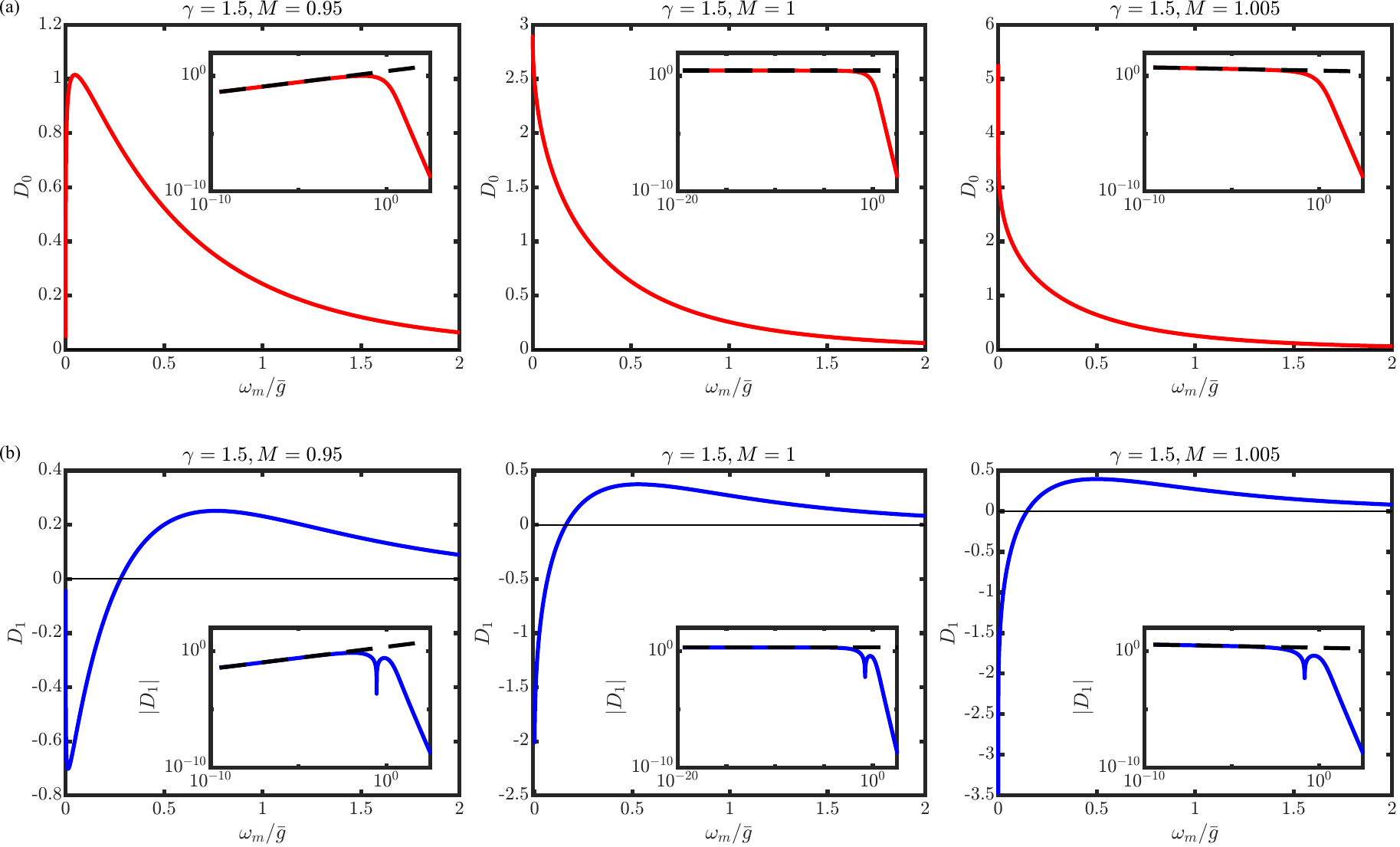}
  \caption{The gap functions $D_0 (\omega_m)$ at $a_\gamma = a_{\gamma,0} (M)$ and $D_1 (\omega_m)$ at $a_\gamma = a_{\gamma,1} (M)$ for $\gamma =1.5$ and $M = 0.95, 1, 1.005$. Insets shows the gap function in logarithmic coordinates, where the black dashed lines indicate the expected power-law behavior
    $\omega^{\alpha_1}_m$, where $\alpha_1$ is the UV exponent from Fig.~\ref{fig:exp}.
    }
  \label{fig:delta_lt2}
\end{figure}

In Fig.~\ref{fig:pd_lt2_v2}, we show the numerical solution of  Eq.~(\ref{3_12b}), obtained at $T \sim 10^{-10} {\bar g}$ for $\gamma =1.5$, which is a representative of $\gamma <2$. Panel (b) on this figure is the expected behavior at $T= 0+$. There is a set of critical lines $a_{\gamma,n} (M)$,which at $T = 0+$ all terminate at $M = M_{cr}$. These lines cross $M=1$ at a set of $a_{\gamma,n}$, the same as in Fig.~\ref{fig:pd_0B}. The observation, most essential to our current analysis, is in Fig.~\ref{fig:delta_lt2}, where we plot the  gap functions $D_n (\omega_m)$ with $n=0,1$ for critical $a_{\gamma,n}$ at some $M < M_{cr}$. Extracting the exponent at small $\omega_m$  at various $M$ (see the insets of Fig.~\ref{fig:delta_lt2}), we find with high degree of accuracy that it is UV exponent $\alpha_1$ from Fig. \ref{fig:exp} {\it for all} $M < M_{cr}$. This holds for both $M <1$, where $\alpha_1 >0$  and for $1< M < M_{cr}$, where $\alpha_1 <0$. We see from Fig. ~\ref{fig:pd_lt2_v2}
 that the solution of the gap equation, subject to the boundary condition $D_n (\omega_m) \propto |\omega_m|^{\alpha_1}$,
exists for
$a_{\gamma}$
 larger than some threshold value. At the threshold, $M = M_{cr}$, and the exponents $\alpha_1$ and $\alpha_2$ merge.
At smaller
$M$
 a non-zero $D_n (\omega_m)$ emerges because the exponents become complex.

   \begin{figure}
  \includegraphics[width=0.9\textwidth]{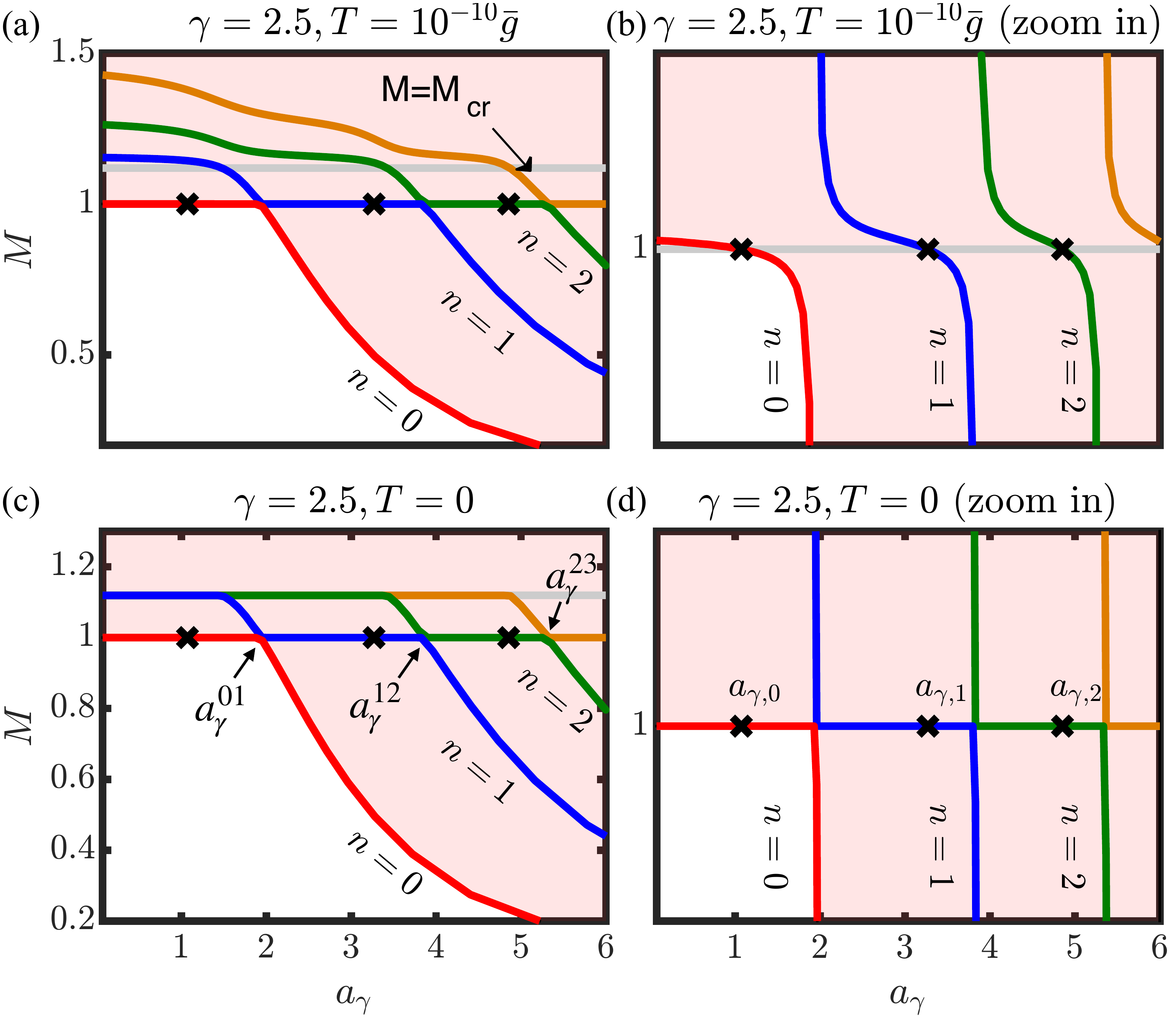}
  \caption{The phase diagram in the $(M, a_\gamma)$ plane for $\gamma =2.5$.
  (a) Numerical solution of Eq. (\ref{3_12b}) at $T =10^{-10}$, $\omega_D=0$; (c) expected result at $T = 0+$, $\omega_D=0$.
  (b) and (d) show zoomed behavior near $M=1$.
  This phase diagram is qualitatively different
   from that for $\gamma <2$. Namely, at a small but finite $T$ the critical line $a_{\gamma_0}$ flattens up at $M \approx 1$ below some
    $a^{01}_\gamma > a_{\gamma,0}$, while the line $a_{\gamma,n}$
     flattens up between $a^{n-1,n}_\gamma < a_{\gamma,n} < a^{n,n+1}_\gamma$ and then continues at $1<M< M_{cr}$ as $a_{\gamma,n-1}$.  At $T=0+$, the flat region coincides with $M=1$. Still, at any $T >0$, the line $a_{\gamma,n}$ crosses $M=1$ at $a_\gamma= a_{\gamma,n}$.  }
  \label{fig:pd_gt2_v2}
\end{figure}

  \begin{figure}
  \includegraphics[width=0.9\textwidth]{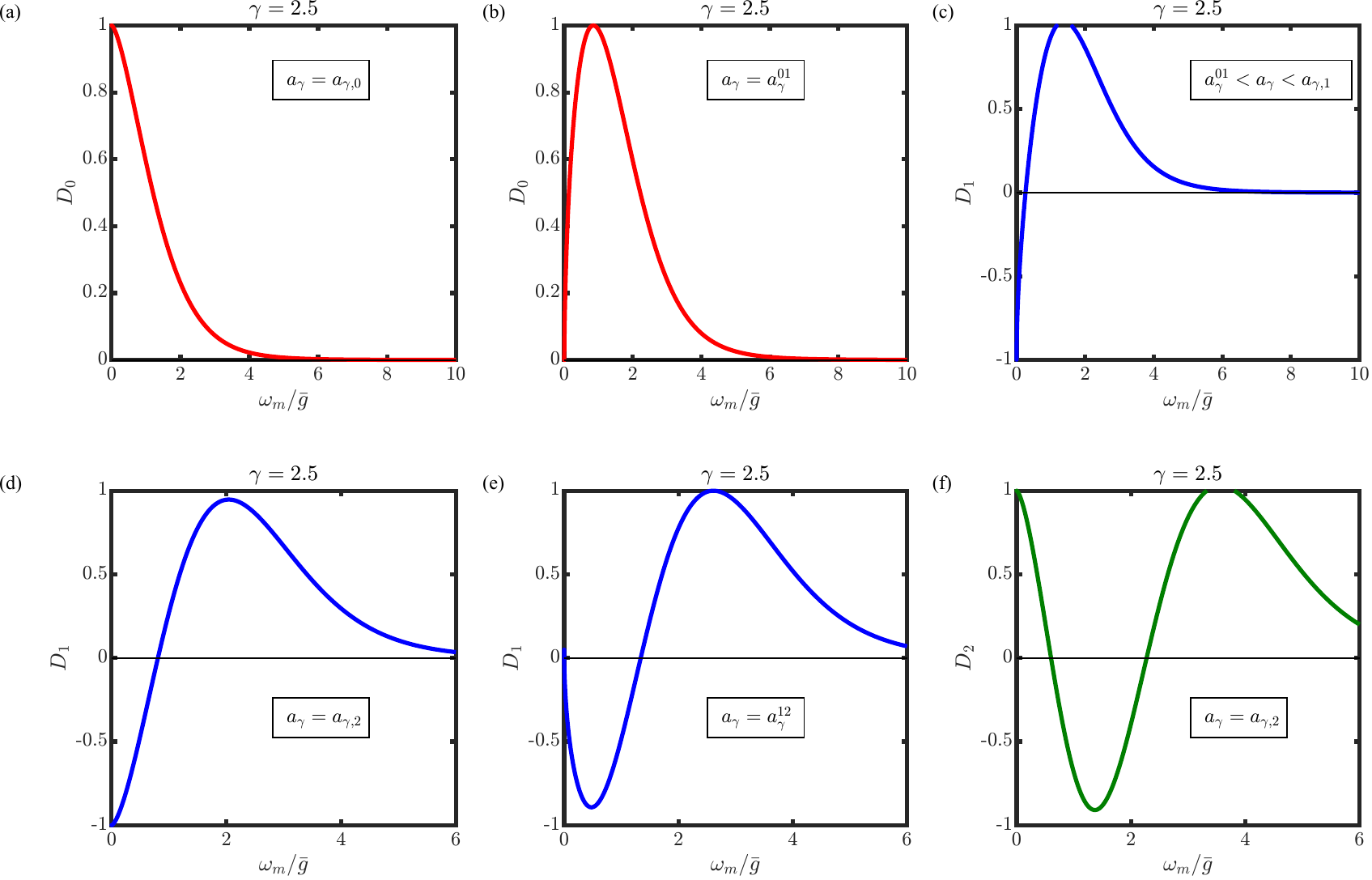}
  \caption{The gap functions $D_0 (\omega_m)$, $D_1 (\omega_m)$ and $D_2 (\omega_m)$ at various values of $a_\gamma$ for $\gamma =2.5$ and $M \approx 1$. The functions $D_0 (\omega_m)$ at  $a_\gamma = a^{01}_{\gamma}$ and $D_1 (\omega_m)$ at $a_\gamma = a^{12}_{\gamma}$ behave at small frequencies as $|\omega_m|^{\gamma-2}$ ($= |\omega_m|^{0.5}$).  At
  $a_\gamma = a_{\gamma,n}$, the corresponding $D_n (\omega_m)$ tend to constant values at $\omega_m =0$ and are analytic at small $\omega_m$.  At intermediate
  $ a^{01}_{\gamma} <a_\gamma < a_{\gamma,1}$, $D_1 (\omega_m)$ behaves at small frequencies as $a +b |\omega_m|^{1/2}$
    }
  \label{fig:delta_gt2}
\end{figure}

Let's now repeat the same calculation for $\gamma >2$.  We show the results in Figs.~\ref{fig:pd_gt2_v2} and \ref{fig:delta_gt2}.
From the numerical solution at small but finite $T$
 (Fig.~\ref{fig:pd_gt2_v2} (a,b)) and its extension to $T=0+$
 (Figs.~\ref{fig:pd_gt2_v2} (c,d)),
we see that while for any $T >0$, the line $a_{\gamma,n}$ crosses $M=1$ at $a_\gamma= a_{\gamma,n}$,
 as in Fig.~\ref{fig:pd_lt2_v2},
  the line $a_{\gamma, 0}$ flattens up at $M \approx 1$ below some
    $a^{01}_\gamma > a_{\gamma,0}$ and the line $a_{\gamma,n}$ with $n>0$
     flattens up between $a^{n-1,n}_\gamma < a_{\gamma,n}$ and $a^{n,n+1}_\gamma > a_{\gamma,n}$ and then continues at $1<M< M_{cr}$ as $a_{\gamma,n-1}$. At $T=0+$, the flat region coincides with $M=1$.

\begin{figure}
  \includegraphics[width=10cm]{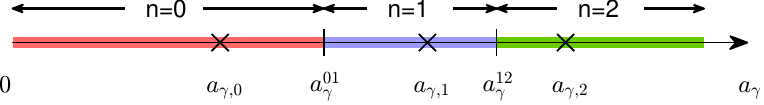}
  \caption{The case $\gamma >2$. The original model with $M=1$ remains critical
 at $T=0+$
   towards pairing in a topological sector specified by $n$ in the interval $a^{n-1,n}_\gamma < a_{\gamma} < a^{n,n+1}_\gamma$.}
  \label{fig:fig2}
\end{figure}

Analyzing the low-frequency behavior of $D_n (\omega_m)$, we find
 that
 for $M >1$ and $M <1$, $D_n (\omega_m) \propto |\omega_m|^{\alpha_1}$,  where $\alpha_1 = \alpha_1 (M)$ is the UV exponent from Fig.~\ref{fig:exp}.
 This exponent is positive for all $M < M_{cr}$, hence $D_n (\omega_m)$ vanishes at $\omega_m =0$.
At $M \to 1$, $\alpha_1 = \gamma -2 >0$. This behavior is different from the one at  $a_\gamma = a_{\gamma,n}$, where
$D_n (0)  = \text{const.} \neq 0$.
 This last behavior is described by the IR exponent $\alpha_2$, which for $\gamma >2$ passes through zero at $M=1$.  Not surprisingly then, the line $a_{\gamma,n} (M)$ reaches $M=1$ at a different value $a_\gamma = a^{n,n+1}_\gamma > a_{\gamma,n}$.  Further, we see from Fig.~\ref{fig:delta_gt2} (c-e) that at
$a_{\gamma}^{01} < a_{\gamma} < a^{12}_\gamma$
 ($M\simeq 1$),
the function  $D_1 (\omega_m)$ at small $\omega_m$ contains both exponents $\alpha_1 = \gamma-2$ and $\alpha_2=0$.  We verified that this holds for other $D_n (\omega_m)$, e,g., at $a_\gamma < a^{01}_\gamma$,
$D_0 (\omega_m) = a^{\prime} + b^{\prime} |\omega_m|^{\gamma-2}$, where $a^{\prime}$ vanishes at $a_\gamma = a^{01}_\gamma$
 (Fig.~\ref{fig:delta_gt2} (b))
and $b^{\prime}$ vanishes at
$a_\gamma = a_{\gamma,1}$
 (Fig.~\ref{fig:delta_gt2} (a)).
This implies that at $T=0+$ (and $\omega_D=0$), the model with $M=1$, which is the one we are interested in,  becomes critical in the sense that the $a_\gamma$ axis gets divided into ranges
 $a^{n-1,n}_\gamma < a_{\gamma} < a^{n,n+1}_\gamma$, where the system remains at the onset of pairing in a topological sector specified by $n$.  We illustrate this in Fig. \ref{fig:fig2}.

\begin{figure}
  \includegraphics[width=0.5\textwidth]{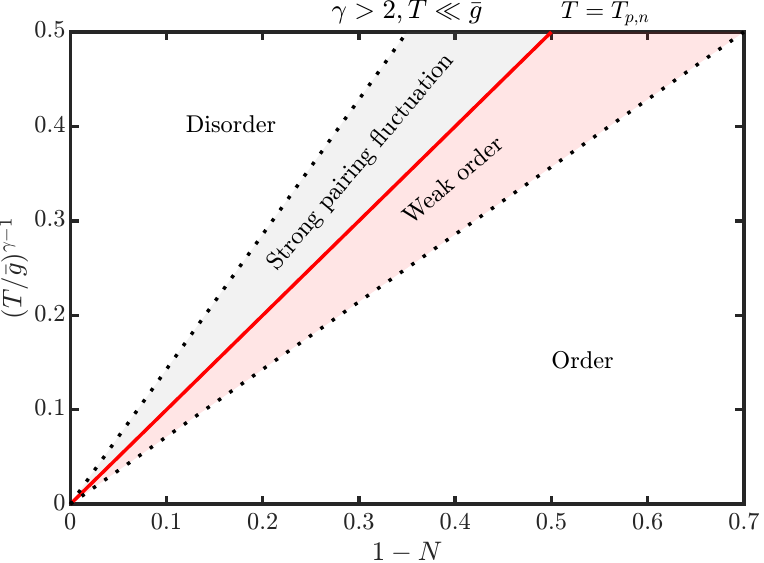}
  \caption{The case $\gamma >2$, $T = 0+$, $1-N=0+$  The region near the critical $T_{p,n} \propto (1-N)^{1/(\gamma-1)}$ for the gap function  $D_n (\omega_m)$  in  the $(T,N)$ plane.
   In one of the two shaded regions around $T_{p,n}$ the system  has near-infinite pairing fluctuations, in the other $D_n (\omega_m)$ is non-zero, but vanishingly  small.  The phase diagram in the $(\omega_D,N)$ plane is quite similar.}
  \label{fig:pd_gt2_v3}
\end{figure}

 At any $T >0$, the flattening at $M=1$ is not exact, and the line $a_{\gamma,n}$ is slightly above $M=1$ for $a_{\gamma} < a_{\gamma,n}$ and slightly below it at  $a_{\gamma,n}  < a_{\gamma} < a^{n,n+1}$
 (see Fig.~\ref{fig:pd_gt2_v2} (b)).
Exactly at  $M=1$ the system then experiences strong pairing fluctuations, but no finite $D_n (\omega_m)$  in the first range, while in the
second range $D_n (\omega_m)$ is non-zero, but very small
(the shaded region in Fig.~\ref{fig:pd_0B}).
We show the phase diagram in the $(T,1-N)$ plane  for a representative $\gamma >2$  in Fig.~\ref{fig:pd_gt2_v3} (the phase diagram in  the $(\omega_D,1-N)$ plane is quite similar).  The two regimes of (i)  near-infinitely strong pairing fluctuations   and (ii) vanishingly small $D_n (\omega_m)$ are in a finite  window around
 $T_{p,n}$.

\begin{figure}
  \includegraphics[width=0.9\textwidth]{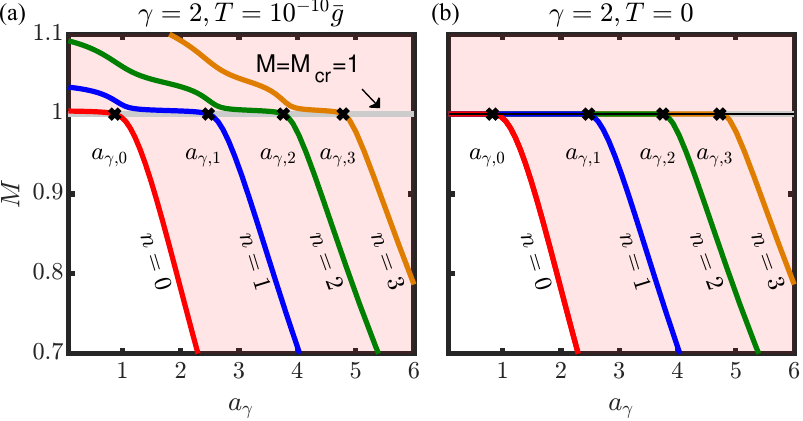}
  \caption{The  phase diagram in the $(M, a_\gamma)$ plane for $\gamma =2$.
   (a) Numerical solution of Eq. (\ref{3_12b}) at $T =10^{-10}$, $\omega_D=0$; (b) expected result at $T = 0+$, $\omega_D=0$. In the last case a critical line $a_{\gamma,n} (M)$ approaches $M=1$ at $a_\gamma = a_{\gamma,n}$ and remains at $M=1$ at smaller $a_\gamma$.}
  \label{fig:pd_eq2_v2}
\end{figure}

  \begin{figure}
  \includegraphics[width=10cm]{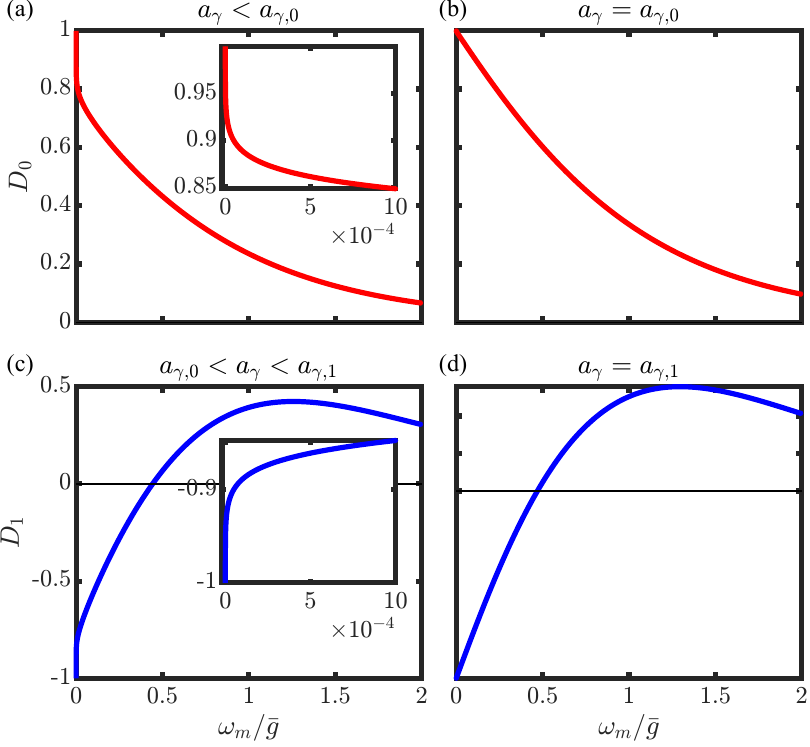}
  \caption{The gap functions $D_0 (\omega_m)$ (panels a,b) and $D_1 (\omega_m)$ (panels c,d) at $\gamma =2$ and $M=1$. Each function scales at low frequencies as $a + b \log({|\omega_m|}/{\bar g})$ for arbitrary $a_{\gamma} < a_{\gamma,n}$. The coefficient $b$ vanishes  at  $a_{\gamma} = a_{\gamma,n}$, where $D_n (\omega_m)$ tends to a constant at $\omega_m =0$ (see panels b,d). Insets indicate the logarithmic behavior at small frequencies.}
  \label{fig:delta_eq2}
\end{figure}

\begin{figure}
  \includegraphics[width=10cm]{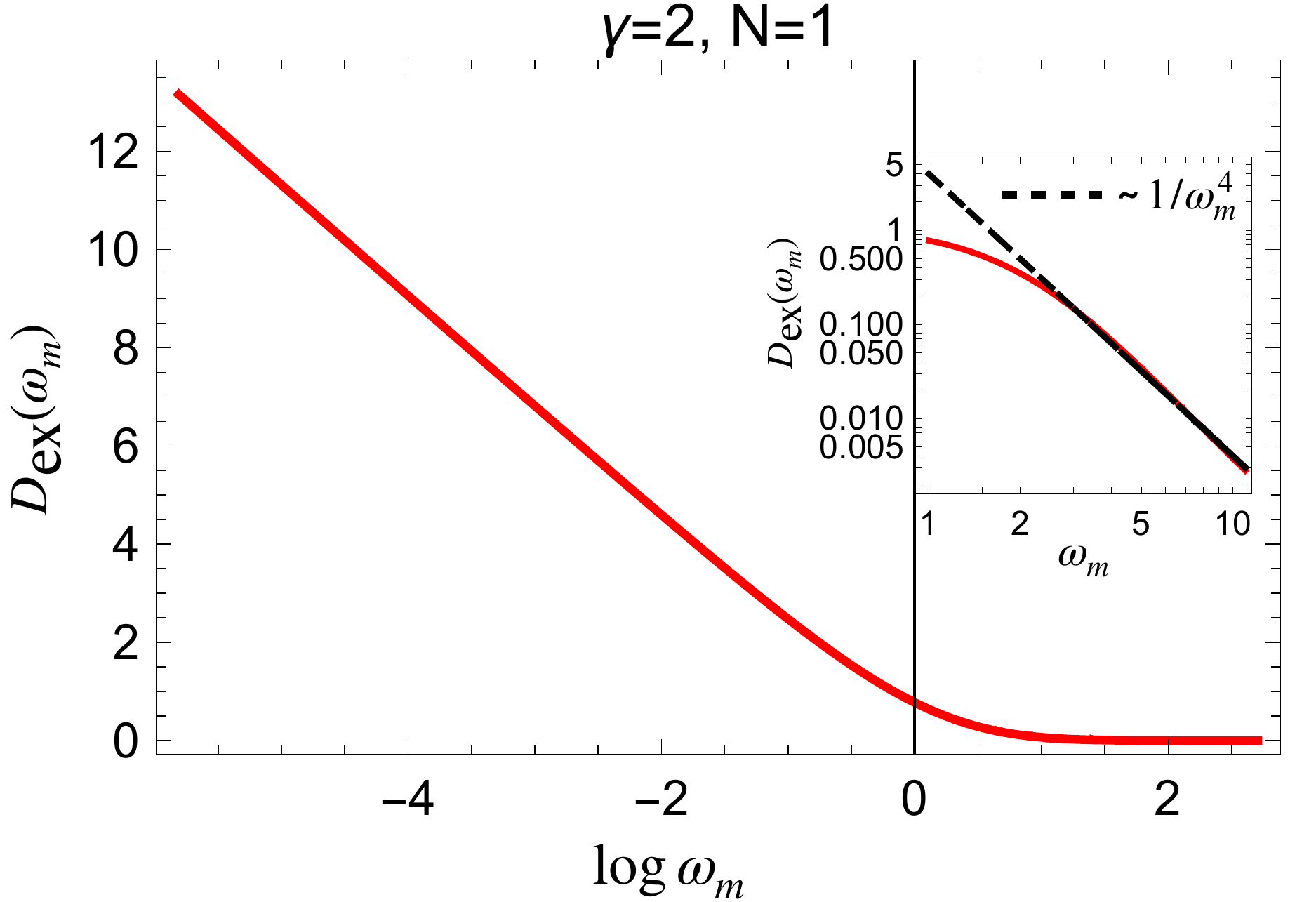}
  \caption{The exact solution of the linearized integral gap equation, $D_{ex}(\omega_m)$,
   for $\gamma=2$ and $M =N=1$ ($a_\gamma =0$).  At small $\omega_m$, $D_{ex}(\omega_m)$ scales as $\log(\omega_m/{\bar g})$, at
     large $\omega_m$ it decays as $1/\omega_m^4$.}\label{fig:exact_g2}
\end{figure}

Finally, the case $\gamma =2$ is the  boundary line between $\gamma <2$ and $\gamma >2$. We show the corresponding behavior  in Figs.~\ref{fig:pd_eq2_v2} and  \ref{fig:delta_eq2}. For $\gamma =2$, $M_{cr} =1$ and $\alpha_1 = \alpha_2=0$ at $M =1$. In this case, $a^{n-1,n}_\gamma = a_{\gamma,n}$. The line $a_{\gamma, n} (M)$ reaches $M=1$ at  $a_\gamma = a_{\gamma,n}$ and remains at $M=1$ at smaller $a_\gamma$.  The function $D_n (\omega_m)$ at $M=1$  scales at low frequencies as $a + b \log({|\omega_m|}/{\bar g})$ for arbitrary $a_{\gamma} < a_{\gamma,n}$. The coefficient $b$ vanishes at  $a_{\gamma} = a_{\gamma,n}$, where $D_n (\omega_m)$ tends to a constant at $\omega_m =0$
 (see Fig.~\ref{fig:delta_eq2} (b,d)).
 At $a_\gamma \to 0$, the frequency range where $D_n (\omega_m)$ changes sign $n$ times shrinks to smaller $\omega_m$ and vanishes at $a_\gamma =0$.
 These topologically distinct solutions $D_n$ merge into the same one $D_{ex} (\omega_m)$ in this limit, which
 can be obtained exactly using the same computational procedure as we used at $N=N_{cr}$ for $\gamma <1$. We show the result for the exact $D_{ex} (\omega_m)$ in Fig. \ref{fig:exact_g2}.

\section{Conclusion and discussion}
\label{sec:conclusion_and_discussion}

In this paper we studied  odd-frequency pairing induced by electron-electron interaction, mediated by a gapless collective boson at a QCP.  The pairing interaction in this case is a function of a frequency transfer, and it  allows for two types of solutions -  even-frequency gap function and odd-frequency gap function.
 We argued that at a QCP, the gap equation has the same form for even-frequency and odd-frequency pairing, despite that
  the spatial structure of the gap function is even vs odd.  We demonstrated this on two examples of pairing at a QCP in 2D:   pairing by Ising-nematic fluctuations and by $(\pi,\pi)$  antiferromagnetic fluctuations. In both cases, we obtained the same gap equation for even-frequency and odd-frequency pairing. The gap equation contains an effective
   dynamical interaction $V (\Omega) = ({\bar g}/|\Omega|)^\gamma$, where $\gamma =1/3$ for a nematic QCP and $1/2$ at an antiferromagnetic QCP. On the Matsubara axis, where the gap function $\Delta (\omega_m)$ can be set as real, the even- and odd-frequency solutions are $\Delta (\omega_m) = \Delta (-\omega_m)$ and $\Delta (\omega_m) = - \Delta (-\omega_m)$, respectively.  For odd-frequency solution, it is more convenient to analyze $D(\omega_m) = \Delta (\omega_m)/\omega_m$ as the latter is even in frequency.

   In our analysis we treated the exponent $\gamma$ as a parameter and analyzed odd-frequency pairing as a function of $\gamma$. The model  with $V (\Omega) \propto 1/|\Omega|)^\gamma$ was
   dubbed the ``$\gamma$-model'', and we used this notation throughout this paper.  We assumed that an even-frequency gap function is suppressed  by an additional frequency-independent  repulsive interaction and
   focused on the odd-frequency solution for the gap.

The key physics of odd-frequency pairing near a QCP is the same as for even-frequency one -- there is a competition between a tendency to pair and a tendency to form a non-Fermi liquid with incoherent quasiparticles.
 Both tendencies originate from the same interaction, which gives rise to pairing when  inserted  in
 to the particle-particle channel, and  to non-Fermi liquid when inserted into the particle-hole channel.  A competition stems from the fact that fermionic incoherence reduces the pairing kernel, while pairing removes spectral weight from  low energies and renders a Fermi liquid behavior.

We  found that in the original $\gamma$-model the tendency towards non-Fermi liquid is stronger for any $\gamma <3$, which we studied, i.e., the ground state is a non-Fermi liquid. However, if the fully dressed interactions in the particle-particle and particle-hole channels are different, and the one in particle-particle channel is larger, the ground state may be an odd-frequency superconductor.  To analyze this, we re-scaled the interaction in the particle-particle channel by a factor $1/N$ and treated $N$ as a model-dependent parameter.  We found that the pairing develops once $N$ is smaller than some  $N_{cr}$.  The value of $N_{cr}$ and the system behavior around this $N$ are different for $\gamma <1$ and $\gamma >1$, which we considered separately.

For $\gamma <1$, we found that $N_{cr}$ depends on $\gamma$. The line
 $N_{cr} (\gamma)$ departs from $N_{cr} (0) =0$  and increases monotonically towards $N_{cr} (1) =1$.
   This implies that the threshold coupling required for odd-frequency pairing, is infinitely large at $\gamma=0$ and becomes
    equal to one at $\gamma=1$.  We analyzed in detail the development of a finite $D(\omega_m)$  at $N \leq N_{cr} (\gamma)$ and found, in similarity to the even-frequency pairing, that  the instability develops when at low frequencies the pairing susceptibility starts oscillating. Mathematically this is caused by the emergence of complex exponents for power-law  form of $D(\omega_m)$ at small frequencies.  Such pairing mechanism is qualitatively different from the BCS one and has a special feature: an
     infinite number of topologically different solutions $D_n (\omega_m)$ emerge simultaneously below $N_{cr}$.
     A topological distinction comes about because $D(\omega_m)$  has $n$ nodal points along the Matsubara frequency axis in the upper half-plane of frequency.  Each  nodal point is the core of a dynamical vortex~\cite{paper_4},
     hence $D_n (\omega_m)$ is the gap function with $n$ vortices.  We found  the sequence of onset temperatures $T_{p,n}$, where $D_n (\omega_m)$ first emerges.

     The $n=0$ solution is sign-preserving and vortex-free. We found that for this solution the onset temperature $T_{p,0}$ is the largest, and the condensation energy at $T=0$ has the largest negative value.  The condensation energy for solutions  with $n >0$ is  smaller by magnitude.
 We analyzed the  forms of the $n=0$ gap function at $T=0$   both on the Matsubara and the real frequency axis.  On the Matsubara axis,  $D_0(\omega_m)$ behaves as $1/|\omega_m|^{\gamma +1}$ at large frequencies and as
 $1/|\omega_m|^d$ at small frequencies, where $0<d<\gamma<1$, i.e., $D_0 (\omega_m)$ diverges at $\omega_m =0$. The exponent $d$ decreases monotonically as $N$ moves towards $N_{cr}$, but tends to a finite value.  On the real axis,
  the implication of this result is that the quasiparticle density of states $N(\omega)$  vanishes at $\omega=0$. We found that $N(\omega)$ scales as $\omega^d$ at small frequencies and has a maximum at $\omega \sim {\bar g}$ (which is the only energy scale in the problem).  The temperature evolution of $N(\omega)$ is rather conventional:   as $T$ increases, the peak frequency  gradually decreases.

   At $\gamma>1$, we found $N_{cr}=1$ independent on $\gamma$.   We  argued that for  any $N <1$, there exists a discrete set of pairing instabilities  at $T_{p,n} \sim (1-N)^{1/(\gamma -1)}/a_{\gamma,n}$, where $a_{\gamma,n}$ increases with $n$.  The same holds at $T=0$ when the bosonic mass $\omega_D$ is non-zero: the instabilities develop at a discrete set of $\omega_D \sim (1-N)^{1/(\gamma -1)}/a_{\gamma,n}$. We obtained this behavior analytically and confirmed numerically that it holds.

We further argued that although the transition lines vary smoothly with $\gamma$, i.e.,  the coefficients $a_{\gamma,n}$ are continuous functions of $\gamma >1$,  there is a qualitative distinction in the low-$T$ behavior at $\gamma<2$ and at $\gamma>2$.  In the latter case, for each $n$ the pairing susceptibility becomes near-infinitely large in a  finite range of $T < T_{p,n}$, and the magnitude of $D_n (\omega_m)$  becomes vanishingly small in a finite range of $T < T_{p,n}$.
 We illustrate this behavior in Fig.~\ref{fig:pd_gt2_v3}.
At $ N=1-0$ and $T = 0+$, this creates a range of $(1-N)/T^{\gamma -1}$, where the system gets frozen at the instability towards pairing in a topological sector specified by $n$.

This last result has interesting implications to field-theory analysis of pairing instabilities out of a non-Fermi liquid as it shows that under some conditions the pairing instability at a QCP does not require complex exponents. We call for more studies to better understand this effect.

\acknowledgements We thank  A. Balatsky, E. Berg, I. Estelis, A. Finkelstein,
A. Klein, D. Maslov, D. Mozyrsky, D. Pimenov, Ph. Phillips, 
V. Pokrovsky, J. Schmalian, A. Tsvelik, and  Y. Wang for useful discussions.  The work by
 Y.M.W., S.-S.Z, and A.V. C.  was supported by the NSF DMR-1834856.
  Y.-M.W, S.-S.Z.,and A.V.C acknowledge the hospitality of KITP at UCSB, where part of the
work has been conducted. The research at KITP is supported by the
National Science Foundation under Grant No. NSF PHY-1748958.

\bibliography{oddfreq}

\end{document}